\documentclass[prd,showpacs,floatfix,amsmath,amssymb,floatfix]{revtex4}
\usepackage{amsmath, amsthm, amssymb, appendix, bm, graphicx,subcaption, mathrsfs}
\usepackage{dcolumn,booktabs,bm}
\usepackage{lscape}
\usepackage{txfonts}
\usepackage{overpic}
\usepackage{multirow}
\usepackage{indentfirst}
\usepackage{feynmf}   
\usepackage{tikz-feynman}
\usepackage{slashed}  
\usepackage{cases}
\usepackage{xcolor}
\usepackage{color,ulem}
\usepackage{float}
\usepackage{braket}
\usepackage[  
colorlinks,
pdfborder=000,
linkcolor=red,
anchorcolor=blue,
citecolor=blue,
urlcolor=blue 
]{hyperref}

\begin{document}

	
	\title{Recoil corrections to pentaquark molecules with an SU(3) anti\--triplet heavy baryon}
	\author{Xiao Chen$^{1}$}\email{xiaochen@bjtu.edu.cn}
	\author{Li Ma$^{1}$}\email{ma.li@bjtu.edu.cn}
	
	\affiliation{
		$^1$School of Physical Science and Engineering, Beijing Jiaotong University, Beijing 100044, China
		}
	
	\date{\today}

	\begin{abstract}	
	  Recoil corrections, which appear at order $\mathcal{O}(\frac{1}{M})$, turn out to be crucial for the pentaquark molecules with heavy flavor. In the past, such corrections were typically regarded as negligible for the formation of heavy molecules. However, our research manifests the important impacts on the spectra of possible baryon-meson bound states. In certain cases, the binding energy sharply decreases after considering the recoil corrections. Using the one-boson-exchange (OBE) model, we have studied a series of double heavy and hidden heavy pentaquark states with an SU(3) anti\--triplet baryon, including $\Xi_{c}\bar{D}^{(*)}$, $\Lambda_{c}\bar{D}^{(*)}$, $\Xi_{c}D^{(*)}$, $\Lambda_{c}D^{(*)}$, along with their bottom analogues. Considering both the $S$\--{}$D$ wave mixing effect and the recoil corrections, we find that recoil corrections, working against the stability of bound states in certain isospin-singlet channels, which cannot be ignored. For the $\Xi_{c}\bar{D}$ and $\Xi_{c}D$ systems with $I(J^P)=0(\frac{1}{2}^{-})$, the $\Xi_{c}\bar{D}^{*}$ and $\Xi_{c}D^{*}$ systems, as well as their bottom counterparts, with $I(J^P)=0(\frac{1}{2}^{-})$ and $0(\frac{3}{2}^{-})$, and the $\Lambda_{b}\bar{B}^{*}$ system with $I(J^P)=\frac{1}{2}(\frac{1}{2}^{-})$ and $\frac{1}{2}(\frac{3}{2}^{-})$, although the bound state solution can be found both with and without considering the recoil corrections, the inclusion of recoil corrections weakens the attraction of the molecules. This phenomenon arising from recoil corrections is especially prominent in the $\Xi_{c}\bar{D}^{*}$ and $\Xi_{c}D^{*}$ systems.   
	\end{abstract}
	
	 \maketitle

	\section{Introduction} \label{sec1}
    Given that quantum chromodynamics (QCD), the first principle of strong interaction, is non-perturbative in the low energy region, physicists have developed various approaches to explore the nature of hadrons. One of the earliest and most influential frameworks is the traditional quark model \cite{ref1,ref2}, in which hadrons are classified into baryons and mesons, consisting of three quarks ($qqq$) and a quark-antiquark pair ($q\bar{q}$), respectively. Although the traditional quark model has successfully described and predicted many experimental observations throughout the 20th century, it is widely believed that exotic states, which do not easily fit into the conventional hadron spectrum, may exist.   
    \par 
    With the advances in technology and the accumulation of experimental data, a series of exotic candidates with heavy flavor including $XYZ$ states \cite{ref3,ref4,ref5,ref6,ref7,ref8}, $P_{c}$ states \cite{ref9,ref10} and $T_{cc}$ states \cite{ref11} have been emerging over the past few years, making their theoretical study a topic of considerable interest. The spectra and dynamical behavior of the observed exotic states have been elucidated through various theoretical interpretations, such as hadronic molecule \cite{ref12,ref13,ref14,ref15}, compact multiquark \cite{ref16,ref17,ref18}, hybrid \cite{ref19,ref20,ref21,ref22} and glueball \cite{ref23,ref24,ref25}. 
    \par 
    Among these, the molecular picture is prominent and widely adopted,  largely because most experimental candidates lie close to the threshold of a hadron pair. For examples, $X(3872)$ and $Y(4260)$ are often interpreted as $D\bar{D}^{*}$ and $D\bar{D}_{1}$ molecular tetraquarks, respectively \cite{ref26,ref27,ref28,ref29,ref30,ref31}. $P_{c}(4440)$ and $P_{c}(4457)$ are commonly regarded as $\Sigma_{c}\bar{D}^{*}$ molecular pentaquarks \cite{ref32}. Moreover, this picture demonstrates powerful predictive capabilities for hadron spectra and decay branching ratios in previous works, which provides valuable guidance for future experimental investigations. In 2013, $DD^{*}$ system with the quantum numbers of $I(J^{P})=0(1^{+})$ was theoretically believed to be a possible molecular state candidate, which was later confirmed by the LHCb observation of $T_{cc}^{+}(3875)$ in 2021 \cite{ref33,ref34}.   
    \par
    Inspired by Yukawa's research on nuclear forces in 1935 \cite{ref35}, the potential model emerged and developed as a powerful tool for describing the effective interactions of hadronic molecules. This approach postulates that the strong interaction between the constituent hadrons can be equivalently treated as being mediated by meson exchanges. Its simplest and earliest version is the one-pion-exchange (OPE) model, which includes only the contribution from $\pi$ exchange. The one-boson-exchange (OBE) model was later proposed to tackle the intermediate and long-range forces, in which the $\pi$, $\eta$, $\sigma$, $\rho$ and $\omega$ exchanges are simultaneously taken into account. Over the following decades, various dynamical effects including the $S$\--$D$ wave mixing effect, the coupled\--channel effect, the isospin breaking effect, and the recoil corrections have been incorporated into the OBE model to promote its accuracy and predictive power. Among them, the recoil corrections are generally perceived insignificant in the heavy flavor systems because they correspond to higher\--order terms suppressed by the large masses of constituent hadrons. Nevertheless, our recent study on the $D_{1}D_{1}$ and the $D_{1}\bar{D}_{1}$ molecules indicates that, in certain cases, the impact of recoil corrections has been severely underestimated, thus calling for more systematic and in-depth investigations \cite{ref36}.   
    \par 
    Recently, the LHCb collaboration reported robust evidence for the $P_{cs}(4338)$ and $P_{cs}(4459)$ states \cite{ref10,ref37}, which has attracted significant theoretical attention. Given that their spectra are close to the threshold of $\Xi_{c}\bar{D}$ and $\Xi_{c}\bar{D}^{*}$, respectively, some previous research investigated them within the hadronic molecular picture \cite{ref38,ref39,ref40,ref41,ref42}. However, several aspects remain unexplored. Within the flavor SU(3) framework, $\Xi_{c}$ and $\Lambda_{c}$ are both belong to the $\bar{\mathbf{3}}$ representation, we naturally extend our studies to the systems containing $\Lambda_{c}$. Furthermore, as the hidden charm molecular states are the possible candidates of $P_{cs}$ states, their open charm counterparts also represent the candidates of $T_{cc}$. Additionally, the bottom partners of all such systems are of equal interest. Consequently, a comprehensive investigation of double heavy and hidden heavy molecular pentaquark states with an anti\--triplet baryon is clearly important.
    \par  
    This work is dedicated to the ground state spectra of possible double heavy and hidden heavy molecular pentaquarks with an anti\--triplet baryon including $\Xi_{c}\bar{D}^{(*)}$, $\Lambda_{c}\bar{D}^{(*)}$, $\Xi_{c}D^{(*)}$, $\Lambda_{c}D^{(*)}$, and their bottom analogues. We employ the OBE model and consider the $S$\--{}$D$ wave mixing effect. Our primary objective is to examine the influence of recoil corrections on the spectra of molecular state solutions. The momentum-related terms are preserved up to $\mathcal{O}(1/M^{2})$, where $M$ denotes the mass of one of the components. In order to elucidate the impacts of recoil corrections, other higher\--order effects, such as the coupled\--channel effect, are not included. This allows us to clearly attribute any spectral changes to the recoil effect itself.
    \par 
    The paper is organized as follows. After introduction, the formalism of the discussed systems will be shown in Sec. \ref{sec2}. With the obtained effective potentials, we will present and analyze the computational results for the mass spectrum of each system in Sec. \ref{sec3}. Finally, we will give the overall discussion and conclusion in Sec. \ref{sec4}. Some intermediate results will be listed in the Appendix.


	\section{Formalism} \label{sec2}
	In the following section, we will construct the ground state wave functions and the relativistic Lagrangian, and derive the effective potentials for the hidden charm systems. Additionally, the derived potentials are then extended to open-charm systems via the $G$-parity rule. Apart from the difference in the heavy flavor hadron masses, the formalism of their bottom analogies is identical.

	\subsection{Wave functions}\label{subsecA} 
	The wave function of a baryon-antimeson system is the product of its isospin, spatial and spin components. The isospin wave function yields an isospin factor in the effective potential, whereas the spatial and wave functions collectively govern the structure of the ground state. 
	\begin{equation}
		\psi^{tot}=\psi^{I} \otimes \psi^{L} \otimes \psi^{S}. \label{1}
	\end{equation}
	\par
	First, we discuss the isospin wave function $\psi^{I}$. The $\Xi_{c}$ baryon and the $D^{(*)}$ meson each carry isospin $I=\frac{1}{2}$, while the $\Lambda_{c}$ baryon is an isospin singlet ($I=0$). Therefore, the $\Xi_{c}\bar{D}^{(*)}$ system can couple to both $I=1$ and $I=0$ channels, while the $\Lambda_{c}\bar{D}^{(*)}$ system is restricted to $I=\frac{1}{2}$ channel. We list the explicit isospin wave functions for these two systems in TABLE \ref{tab:1} and TABLE \ref{tab:2}, respectively. Due to isospin conservation in the strong interaction, the effective potentials only depend on the total isospin $I$, and are independent of its third component $I_{3}$.
	\par
	Then, we consider the spatial and the spin wave functions together. Due to the centrifugal barrier, a bound state solution is more likely to appear in the ground states, and that is why the higher partial waves are excluded in this research. The ground state of the systems without tensor force interaction is the pure $S$ wave, whereas the ground state of the systems with tensor force interaction is the superposition of the $S$ and the $D$ wave for such interaction leads to the $S$\--{}$D$ wave mixing. 
	\par 
	For the $\Xi_{c}\bar{D}$ and $\Lambda_{c}\bar{D}$ systems, the total angular momentum only couple to $J=\frac{1}{2}$, and the ground state can be written as
	\begin{eqnarray}
		\ket{ \psi ^{J=\frac{1}{2}} } =\ket{ {^2}S_{\frac{1}{2}}}. \label{2}
	\end{eqnarray}
	\par 
	Given that $\bar{D}^{*}$ is a vector anti-meson, both $J=\frac{1}{2}$ and $J=\frac{3}{2}$ channels are possible for the $\Xi_{c}\bar{D}^{*}$ and $\Lambda_{c}\bar{D}^{*}$ systems, and the ground states can be written as
	\begin{align}
		\ket{ \psi ^{J=\frac{1}{2}}}  &=R_s\ket{ {^2}S_{\frac{1}{2}}} +R_d \ket{ {^4}D_{\frac{1}{2}}}, \label{3}\\ 
		\ket{ \psi ^{J=\frac{3}{2}} } &=R_{s1}\ket{ {^4}S_{\frac{3}{2}}} +R_{d1} \ket{ {^2}D_{\frac{3}{2}}}+R_{d2} \ket{ {^4}D_{\frac{3}{2}}}. \label{4}
	\end{align}
	\par
	The $D$ wave coefficients $R_{d}$, $R_{d1}$, and $R_{d2}$ are nonzero only when recoil corrections are included. This is because the tensor force related terms emerge in the effective potentials only if the momentum-related terms are considered in our cases.
	\par  
	Based on the above analysis, all possible channels of the hidden charm systems are summarized as follows. For the $\Xi_{c}\bar{D}$ system, channels with $I(J^P)=0(\frac{1}{2}^{-}), 1(\frac{1}{2}^{-})$ are allowed; for the $\Lambda_{c}\bar{D}$ system, only the channel with $I(J^P)=\frac{1}{2}(\frac{1}{2}^{-})$ is accessible; for the $\Xi_{c}\bar{D}^{*}$ system, channels with $I(J^P)=0(\frac{1}{2}^{-}), 0(\frac{3}{2}^{-}), 1(\frac{1}{2}^{-})$, $1(\frac{3}{2}^{-})$ are possible; for the $\Lambda_{c}\bar{D}^{*}$ system, channels with $I(J^P)=\frac{1}{2}(\frac{1}{2}^{-})$, $\frac{1}{2}(\frac{3}{2}^{-})$  are supported.
	
		\renewcommand{\arraystretch}{1.5}
	\begin{table*}[!htb]
		\scriptsize
		\begin{center}
			\caption{\label{tab:1} The isospin wave functions for the $\Xi_{c}\bar{D}^{(*)}$ system}
			\begin{tabular}{c|c}\toprule[1pt]
				$\left|I, I_{3}\right\rangle$ & $\Xi_{c}\bar{D}^{(*)}$ \text { system }  \\
				\midrule[1pt]
				$|1,1\rangle$ & $\left | \Xi_{c}^{+}\bar{D}^{(*)0}\right\rangle$  \\
				$|1,0\rangle$ & $\frac{1}{\sqrt{2}}\left(\left|\Xi_{c}^{+}D^{(*)-}\right\rangle+\left |\Xi_{c}^{0}\bar{D}^{(*)0}\right\rangle\right)$  \\
				$|1,-1\rangle$ & $\left|\Xi_{c}^{0}D^{(*)-}\right\rangle$  \\
				$|0,0\rangle$ & $\frac{1}{\sqrt{2}}\left(\left|\Xi_{c}^{+}D^{(*)-}\right\rangle-\left|\Xi_{c}^{0}\bar{D}^{(*)0}\right\rangle\right)$ \\
				
				\bottomrule[1pt]
				
			\end{tabular}
		\end{center}
	\end{table*}
	
	\renewcommand{\arraystretch}{1.5}
	\begin{table*}[!htb]
		\scriptsize
		\begin{center}
			\caption{\label{tab:2} The isospin wave functions for the $\Lambda_{c}\bar{D}^{(*)}$ system}
			\begin{tabular}{c|c}\toprule[1pt]
				$\left|I, I_{3}\right\rangle$ & $\Lambda_{c}\bar{D}^{(*)}$ \text { system }  \\
				\midrule[1pt]
				$|\frac{1}{2},\frac{1}{2}\rangle$ & $\left | \Lambda_{c}^{+}\bar{D}^{(*)0}\right\rangle$  \\
				$|\frac{1}{2},-\frac{1}{2}\rangle$ & $\left | \Lambda_{c}^{+}\bar{D}^{(*)-}\right\rangle$  \\
				
				\bottomrule[1pt]
				
			\end{tabular}
		\end{center}
	\end{table*}

	\subsection{Relativistic effective Lagrangians and Effective potentials}\label{subsecB}
	The relativistic effective Lagrangians are supposed to include all possible structures satisfying the symmetries of the strong interaction---namely, parity, charge conjugation, and isospin conservation. Besides, they should degenerate to its non-relativistic form after applying the heavy quark limit. Therefore, the coupling constants in this work are extracted from previous research that did not consider the recoil corrections. Within the OBE model, pseudoscalar, scalar and vector meson exchanges contribute to the interaction. However, previous phenomenological analyses have shown that the coupling constants of the vertices between baryons in the $\bar{\mathbf{3}}$ representation and the pseudoscalar mesons are equal to zero \cite{ref43}. 
	\par 
	We introduce the following notations to represent the constituent hadrons and the exchanged light mesons, which read
	\begin{equation}
		\begin{aligned}
			\Lambda_{c} = \Lambda_{c}^{+}, \quad 
			\Xi_{c} = \binom{\Xi_{c}^{+}}{\Xi_{c}^{0}}, \quad  
			\bar{D}^{(*)} = \binom{\bar{D}^{(*)0}}{D^{(*)-}}, \\
			\sigma = \sigma, \quad 
			V_{\mu} = 
			\begin{pmatrix}
				\frac{\rho^{0}}{\sqrt{2}} + \frac{\omega}{\sqrt{2}} & \rho^{+} \\
				\rho^{-} & -\frac{\rho^{0}}{\sqrt{2}} + \frac{\omega}{\sqrt{2}}
			\end{pmatrix}_{\mu}.
		\end{aligned}
	\end{equation}
	\par 
	The Lagrangians for interactions between $\bar{\mathbf{3}}$ baryons and the exchanged light mesons read 
	    \begin{align}
	    	\mathcal{L}^{R}_{\Xi_c \Xi_c \sigma }&=g_{\Xi_c \Xi_c \sigma }\bar{\Xi}_c \sigma  \Xi_c, \\
	    	\mathcal{L}^{R}_{\Xi_c \Xi_c V }&=g_{\Xi_{c}\Xi_{c} V}\left(\bar{\Xi}_{c}\gamma^{\mu} \Xi_{c}\right)V_{\mu}+\frac{1}{4M_{\Xi_{c}}}f_{\Xi_{c}\Xi_{c}V}\left(\bar{\Xi}_{c}\sigma^{\mu\nu}\Xi_{c}\right)\left(\partial_\mu V_\nu -\partial_\nu V_{\mu}\right),\\
	    	\mathcal{L}^{R}_{\Lambda_c \Lambda_c \sigma }&=g_{\Lambda_c \Lambda_c \sigma }\bar{\Lambda}_c \sigma  \Lambda_c, \\
	    	\mathcal{L}^{R}_{\Lambda_c \Lambda_c V }&=g_{\Lambda_{c}\Lambda_{c} V}\left(\bar{\Lambda}_{c}\gamma^{\mu} \Lambda_{c}\right)V_{\mu}+\frac{1}{4M_{\Lambda_{c}}}f_{\Lambda_{c}\Lambda_{c}V}\left(\bar{\Lambda}_{c}\sigma^{\mu\nu}\Lambda_{c}\right)\left(\partial_\mu V_\nu -\partial_\nu V_{\mu}\right).
	    \end{align}
	    \par 
	    The Lagrangians for interactions between $\bar{D}^{(*)}$ mesons and the exchanged light mesons read
	    \begin{align}
	    	\mathcal{L}^{R}_{\bar{D}\bar{D}\sigma}&=-2g_{\sigma}M_D \bar{D}\sigma\bar{D}^{\dagger},\\
	    	\mathcal{L}^{R}_{\bar{D}\bar{D} V}&=-i\frac{\sqrt{2}}{2}\beta g_{V}\left(\bar{D}\overleftrightarrow{\partial^{\mu}}\bar{D}^{\dagger}\right)V_{\mu},\\
	    	\mathcal{L}^{R}_{\bar{D}^{*}\bar{D}^{*}\sigma}&=2g_{\sigma}M_{D^{*}} \bar{D}^{*\mu}\sigma\bar{D}^{*\dagger}_{\mu},\\
	    	\mathcal{L}^{R}_{\bar{D}^{*}\bar{D}^{*} V}&=i\frac{\sqrt{2}}{2}\beta g_{V}\left(\bar{D}^{*\nu}\overleftrightarrow{\partial^{\mu}}\bar{D}^{*\dagger}_{\nu}\right)V_{\mu}-i2\sqrt{2}\lambda g_{V} M_{D^{*}}\bar{D}^{*\mu}\left(\partial_{\mu}V_{\nu}-\partial_{\nu}V_{\mu}\right)\bar{D}^{*\nu \dagger}. 
	    \end{align}	 
		\par
	    In TABLE \ref{tab:3} , we collect the numerical values of the coupling constants and the masses of the hadrons used in our calculations. 
	    \par 
	    Each constituent meson and baryon provides a polarization vector $\epsilon$ and a spinor $u$, respectively. In the rest frame, they read
	    \begin{align}\label{eq:14}
	    	\epsilon=(0,\boldsymbol{\epsilon}), \qquad u=\sqrt{M} (\xi,0 ). 
	    \end{align}
	    To derive the polarization vector $\epsilon$ and the spinor $u$ in the laboratory frame, we make a boost to Eq. (\ref{eq:14}), and the results read
	    \begin{align}
	    	\epsilon ^{Lab}=\left(\frac{\boldsymbol{p}\cdot \boldsymbol{\epsilon }}{m}, \boldsymbol{\epsilon }+\frac{\boldsymbol{p}(\boldsymbol{p}\cdot \boldsymbol{\epsilon })}{m(p_0+m)}  \right), \qquad u^{Lab}=\sqrt{E+M}\left(\xi, \frac{\boldsymbol{\sigma} \cdot \boldsymbol{p}}{E+M}\xi\right)
	    \end{align}     
	\begin{table*}[!htb]
		\scriptsize
		\centering
		\captionsetup{format=plain, justification=raggedright, singlelinecheck=false}
		\caption{\label{tab:3} The masses of the hadrons and the coupling constants used in this study. The masses are extracted from the Particle Data Group's results \cite{ref44}}
		\renewcommand{\arraystretch}{1.2}
		\setlength{\tabcolsep}{4pt}
		\begin{tabular}{@{} l c| l @{}}
			\toprule[1pt]
			& Mass (MeV) & Coupling constants \\
			\midrule[1pt]
			Scalar & $m_\sigma=600$ & $g_{\Xi_{c}\Xi_{c}\sigma}(g_{\Lambda_{c}\Lambda_{c}\sigma})=-3.65$ \cite{ref38} \\
			& & $g_{\Xi_{c}\Xi_{c}V}(g_{\Lambda_{c}\Lambda_{c}V})=4.59$ \cite{ref45} \\
			Vector & $m_\rho=775.26$ & $f_{\Xi_{c}\Xi_{c}V}(f_{\Lambda_{c}\Lambda_{c}V})=-4.59$ \cite{ref45} \\
			& $m_\omega=782.66$ & $g_{\sigma}=0.761$ \cite{ref46}\\ 
			& & $g_{V}=5.83$ \cite{ref47} \\
			\midrule
			Heavy flavor & $m_{D}=1864.8$ & $\beta=0.9$ \cite{ref48} \\
			& $m_{D^{*}}=2006.9$ & $\lambda=-0.56$ GeV$^{-1}$ \cite{ref47}\\
			& $m_{B}=5279.7$ & \\
			& $m_{B^{*}}=5324.8$ & \\
			& $m_{\Xi_{c}}=2470.4$ & \\
			& $m_{\Xi_{b}}=5791.9$ & \\       
			& $m_{\Lambda_{c}}=2286.5$ & \\
			& $m_{\Lambda_{b}}=5619.6$ & \\
			\bottomrule[1pt]
		\end{tabular}
	\end{table*}
    \par 
    Using the effective Lagrangians, we extract the Feynman rules for the vertices and calculate the corresponding scattering amplitudes. The effective potentials are subsequently derived in the momentum representation, and then they are transformed into the coordinate space.
    \begin{center}
    	\begin{minipage}{0.24\textwidth}
    		\centering
    		\begin{tikzpicture}
    			\begin{feynman}		
    				\vertex (a) ;
    				\vertex [right=2.5 of a](b);
    				\vertex [above=1.7 of a](f3) {$\Xi_{c}$};
    				\vertex [below=1.7 of a](f1) {$\Xi_{c}$};
    				\vertex [above=1.7 of b](f4) {$\bar{D}$};
    				\vertex [below=1.7 of b](f2) {$\bar{D}$};
    				\diagram* {
    					(b) -- [edge label'=$\sigma \quad \rho \quad \omega$,momentum=$q$] (a) -- [fermion, edge, momentum=$P_3$] (f3),
    					(f1) -- [fermion, edge, momentum=$P_1$] (a), 
    					(b),
    					(f2) -- [scalar, momentum=$P_2$] (b) --[scalar, momentum=$P_4$] (f4),
    				};
    			\end{feynman} 	  
    		\end{tikzpicture}
    	\end{minipage}
    	\hfill
    	\begin{minipage}{0.24\textwidth}
    		\centering
    		\begin{tikzpicture}
    			\begin{feynman}		
    				\vertex (a) ;
    				\vertex [right=2.5 of a](b);
    				\vertex [above=1.7 of a](f3) {$\Xi_{c}$};
    				\vertex [below=1.7 of a](f1) {$\Xi_{c}$};
    				\vertex [above=1.7 of b](f4) {$\bar{D}^{*}$};
    				\vertex [below=1.7 of b](f2) {$\bar{D}^{*}$};
    				\diagram* {
    					(b) -- [edge label'=$\sigma \quad \rho \quad \omega$,momentum=$q$] (a) -- [fermion, edge, momentum=$P_3$] (f3),
    					(f1) -- [fermion, edge, momentum=$P_1$] (a), 
    					(b),
    					(f2) -- [boson, momentum=$P_2$] (b) --[boson, momentum=$P_4$] (f4),
    				};
    			\end{feynman} 	  
    		\end{tikzpicture}
    	\end{minipage}
    	\begin{minipage}{0.24\textwidth}
    		\centering
    		\begin{tikzpicture}
    			\begin{feynman}		
    				\vertex (a) ;
    				\vertex [right=2.5 of a](b);
    				\vertex [above=1.7 of a](f3) {$\Lambda_{c}$};
    				\vertex [below=1.7 of a](f1) {$\Lambda_{c}$};
    				\vertex [above=1.7 of b](f4) {$\bar{D}$};
    				\vertex [below=1.7 of b](f2) {$\bar{D}$};
    				\diagram* {
    					(b) -- [edge label'=$\sigma \quad \omega$,momentum=$q$] (a) -- [fermion, edge, momentum=$P_3$] (f3),
    					(f1) -- [fermion, edge, momentum=$P_1$] (a), 
    					(b),
    					(f2) -- [scalar, momentum=$P_2$] (b) --[scalar, momentum=$P_4$] (f4),
    				};
    			\end{feynman} 	  
    		\end{tikzpicture}
    	\end{minipage}
    	\hfill
    	\begin{minipage}{0.24\textwidth}
    		\centering
    		\begin{tikzpicture}
    			\begin{feynman}		
    				\vertex (a) ;
    				\vertex [right=2.5 of a](b);
    				\vertex [above=1.7 of a](f3) {$\Lambda_{c}$};
    				\vertex [below=1.7 of a](f1) {$\Lambda_{c}$};
    				\vertex [above=1.7 of b](f4) {$\bar{D}^{*}$};
    				\vertex [below=1.7 of b](f2) {$\bar{D}^{*}$};
    				\diagram* {
    					(b) -- [edge label'=$\sigma \quad \omega$,momentum=$q$] (a) -- [fermion, edge, momentum=$P_3$] (f3),
    					(f1) -- [fermion, edge, momentum=$P_1$] (a), 
    					(b),
    					(f2) -- [boson, momentum=$P_2$] (b) --[boson, momentum=$P_4$] (f4),
    				};
    			\end{feynman} 	  
    		\end{tikzpicture}
    	\end{minipage}
    \end{center}
    \par 
    The scattering processes for the hidden charm systems are depicted schematically in the Feynman diagrams above. We firstly define the inter momentum $\boldsymbol{q}$ and the external momentum $\boldsymbol{k}$ 
    \begin{align}
    	\boldsymbol{q}= \boldsymbol{P}_{3}-\boldsymbol{P}_{1}, \qquad
    	\boldsymbol{k}= \frac{1}{2}(\boldsymbol{P}_{3}+\boldsymbol{P}_{1}),
    \end{align} 
    which represent the motion of the exchanged mesons and the constituent hadrons separately.     
    \par 
    Applying the first Born approximation, the effective potential  $V(\boldsymbol{q},\boldsymbol{k})$ can be obtained from the invariant scattering amplitude
    \begin{align}
    	V(\boldsymbol{q},\boldsymbol{k} )=T_{fi}=-\frac{\mathcal{M} }{\sqrt{2E_{1}}\sqrt{2E_{2}}\sqrt{2E_{3}}\sqrt{2E_{4}} },
    \end{align} 
    where $E_{1}$\--$E_{4}$ are energies of the initial and final state hadrons.
    \par By means of the Fourier transformation, we can derive the effective potential in the coordinate space
    \begin{align}\label{eq:18}
    	V(\boldsymbol{r},\boldsymbol{r}^{\prime},\boldsymbol{\nabla _{r}},\boldsymbol{\nabla} _{\boldsymbol{r}^{\prime}})=\frac{1}{(2\pi)^3} \iint \mathrm{d}^3\boldsymbol{q}\mathrm{d}^3\boldsymbol{k}\mathcal{F} (q)^2 V(\boldsymbol{q},\boldsymbol{k} ) e^{i\boldsymbol{k} \cdot (\boldsymbol{r}^{\prime}-\boldsymbol{r})+i\frac{1}{2}\boldsymbol{q} \cdot (\boldsymbol{r}^{\prime}+\boldsymbol{r})},
    \end{align}
    where $\mathcal{F} (q)$ is the form factor to describe the non-point-like
    structures attached to each vertex. With the form factor, the ultraviolet divergence can be well controlled. Here, we apply the monopole form factor
    \begin{align}
    	 \mathcal{F} (q)=\frac{\Lambda^2-m^2 }{\Lambda^2-q^2},
    \end{align}
    given that it achieved an exciting success in explaining the spectrum of deuteron. $\Lambda$ here is an input cutoff parameter.
    \par 
    The effective potentials for the $\Xi_{c}\bar{D}$ and  the$\Xi_{c}\bar{D}^{*}$ systems are given below. The effective potentials share the similar form in the $\Lambda_{c}\bar{D}$ and the $\Lambda_{c}\bar{D}^{*}$ systems except for the baryon masses and the isospin factors. We expand the potentials to $\frac{1}{M^2}$ order, where $M$ represents the masses of the constituent hadrons. The recoil related terms initially emerge at the $\frac{1}{M}$ order. The results of the Fourier transformations in Eq. (\ref{eq:18}) are listed in the Appendix.
  
    \par
    For the $\Xi_{c}\bar{D}$ system, the $\sigma$ exchange potential reads  
    \begin{align}\label{eq:21}   	V^{Lab[\Xi_{c}\bar{D}]}_{scalar}\left(\boldsymbol{r},\boldsymbol{\nabla}\right)
    	=&\mathcal{C}_{\sigma }\left[\frac{m_{\sigma}}{4\pi}g_{\Xi_{c}\Xi_{c}\sigma}g_{\sigma}H_{0}
    	-\frac{m_{\sigma}^{3}}{16\pi M_{D}^2}g_{\Xi_{c}\Xi_{c}\sigma}g_{\sigma}H_{1}
    	-\frac{m_{\sigma}^{3}}{32\pi M_{\Xi_{c}}^2}g_{\Xi_{c}\Xi_{c}\sigma}g_{\sigma}H_{1}\right.\nonumber\\
    	&\left.+\frac{m_{\sigma}^{3}}{8\pi M_{\Xi_{c}}^2}g_{\Xi_{c}\Xi_{c}\sigma}g_{\sigma}H_{2}\left(\boldsymbol{r}\cdot \boldsymbol{\nabla}\right)
    	+\frac{m_{\sigma}^{3}}{8\pi M_{D}^2}g_{\Xi_{c}\Xi_{c}\sigma}g_{\sigma}H_{2}\left(\boldsymbol{r}\cdot \boldsymbol{\nabla}\right)\right.\nonumber\\
    	&\left.+\frac{m_{\sigma}}{8\pi M_{\Xi_{c}}^{2}}g_{\Xi_{c}\Xi_{c}\sigma}g_{\sigma}H_{0}\boldsymbol{\nabla}^{2}
    	+\frac{m_{\sigma}}{8\pi M_{D}^{2}}g_{\Xi_{c}\Xi_{c}\sigma}g_{\sigma}H_{0}\boldsymbol{\nabla}^{2}
    	+\frac{m_{\sigma}^{3}}{16\pi M_{\Xi_{c}}^{2}}g_{\Xi_{c}\Xi_{c}\sigma}g_{\sigma}H_{2}\left(\boldsymbol{\sigma}\cdot\boldsymbol{L}\right)\right],
    \end{align}
    while the vector ($\rho$,$\omega$) exchange potential reads
    \begin{align}\label{eq:22} 	V^{Lab[\Xi_{c}\bar{D}]}_{vector}\left(\boldsymbol{r},\boldsymbol{\nabla}\right)
    	=&
    	\sum_{\alpha}\mathcal{C}_{\alpha}
    	\left[\frac{\sqrt{2}m_{v}}{8\pi}\beta g_{V}g_{\Xi_{c}\Xi_{c}V}H_{0}
    	-\frac{\sqrt{2}m_{v}^{3}}{64\pi M_{\Xi_{c}}^{2}}\beta g_{V}g_{\Xi_{c}\Xi_{c}V}H_{1}
    	+\frac{\sqrt{2}m_{v}^{3}}{32\pi M_{\Xi_{c}} M_{D}}\beta g_{V}g_{\Xi_{c}\Xi_{c}V}H_{1}\right.\nonumber\\
    	&
    	\left.-\frac{\sqrt{2}m_{v}^{3}}{32\pi M_{\Xi_{c}}^{2}}\beta g_{V}f_{\Xi_{c}\Xi_{c}V}H_{1}
    	-\frac{\sqrt{2}m_{v}^{3}}{8\pi M_{\Xi_{c}} M_{D}}\beta g_{V}g_{\Xi_{c}\Xi_{c}V}H_{2}\left(\boldsymbol{r}\cdot \boldsymbol{\nabla}\right)
    	-\frac{\sqrt{2}m_{v}}{8\pi M_{\Xi_{c}} M_{D}}\beta g_{V}g_{\Xi_{c}\Xi_{c}V}H_{0}\boldsymbol{\nabla}^2\right.\nonumber\\
    	&
    	\left.-\frac{\sqrt{2}m_{v}^{3}}{32\pi M_{\Xi_{c}^{2}}}\beta g_{V}g_{\Xi_{c}\Xi_{c}V}H_{2}\left(\boldsymbol{\sigma}\cdot \boldsymbol{L}\right)
    	-\frac{\sqrt{2}m_{v}^{3}}{16\pi M_{\Xi_{c}} M_{D}}\beta g_{V}g_{\Xi_{c}\Xi_{c}V}H_{2}\left(\boldsymbol{\sigma}\cdot \boldsymbol{L}\right) \right.\nonumber\\
    	&
    	\left.-\frac{\sqrt{2}m_{v}^{3}}{16\pi M_{\Xi_{c}}^{2}}\beta g_{V}f_{\Xi_{c}\Xi_{c}V}H_{2}\left(\boldsymbol{\sigma}\cdot \boldsymbol{L}\right)  
    	-\frac{\sqrt{2}m_{v}^{3}}{16\pi M_{\Xi_{c}}M_{D}}\beta g_{V}f_{\Xi_{c}\Xi_{c}V}H_{2}\left(\boldsymbol{\sigma}\cdot \boldsymbol{L}\right)\right].
    \end{align}
    \par 
    For the $\Xi_{c}\bar{D}^{*}$ system, the $\sigma$ exchange potential reads
    \begin{align}\label{eq:23} 	V^{Lab[\Xi_{c}\bar{D}^{*}]}_{scalar}\left(\boldsymbol{r},\boldsymbol{\nabla}\right)=
    	&\mathcal{C}_{\sigma}
    	\left[ \frac{m_{\sigma}}{4\pi}g_{\Xi_{c}\Xi_{c}\sigma}g_{\sigma}H_{0}\left(\boldsymbol{\epsilon}_{2}\cdot\boldsymbol{\epsilon}_{4}^{\dagger}\right)
    	-\frac{m_{\sigma}^{3}}{48\pi M_{D^{*}}^{2}}g_{\Xi_{c}\Xi_{c}\sigma}g_{\sigma}H_{1}\left(\boldsymbol{\epsilon}_{2}\cdot\boldsymbol{\epsilon}_{4}^{\dagger}\right)\right.\nonumber\\
    	&
    	\left.+\frac{m_{\sigma}}{8\pi M_{D^{*}}^{2}}g_{\Xi_{c}\Xi_{c}\sigma}g_{\sigma}H_{0}\left(\boldsymbol{\epsilon}_{2}\cdot\boldsymbol{\epsilon}_{4}^{\dagger}\right)\boldsymbol{\nabla}^2
    	+\frac{m_{\sigma}}{8\pi M_{\Xi_{c}}^{2}}g_{\Xi_{c}\Xi_{c}\sigma}g_{\sigma}H_{0}\left(\boldsymbol{\epsilon}_{2}\cdot\boldsymbol{\epsilon}_{4}^{\dagger}\right)\boldsymbol{\nabla}^2\right.\nonumber\\
    	&
    	\left.-\frac{m_{\sigma}^{3}}{32\pi M_{\Xi_{c}}^{2}}g_{\Xi_{c}\Xi_{c}\sigma}g_{\sigma}H_{1}\left(\boldsymbol{\epsilon}_{2}\cdot\boldsymbol{\epsilon}_{4}^{\dagger}\right)
    	+\frac{m_{\sigma}^{3}}{8\pi M_{D^{*}}^{2}}g_{\Xi_{c}\Xi_{c}\sigma}g_{\sigma}H_{2}\left(\boldsymbol{\epsilon}_{2}\cdot\boldsymbol{\epsilon}_{4}^{\dagger}\right)\left(\boldsymbol{r}\cdot \boldsymbol{\nabla}\right)\right.\nonumber\\
    	&
    	\left.
    	+\frac{m_{\sigma}^{3}}{8\pi M_{\Xi_{c}}^{2}}g_{\Xi_{c}\Xi_{c}\sigma}g_{\sigma}H_{2}\left(\boldsymbol{\epsilon}_{2}\cdot\boldsymbol{\epsilon}_{4}^{\dagger}\right)\left(\boldsymbol{r}\cdot \boldsymbol{\nabla}\right)
    	-\frac{m_{\sigma}^3}{24\pi M_{D^{*}}^{2}}g_{\Xi_{c}\Xi_{c}\sigma}g_{\sigma}H_{3}S(\hat{\boldsymbol{r}},\boldsymbol{\epsilon}_{2},\boldsymbol{\epsilon}_{4}^{\dagger})
    	\right.\nonumber\\
    	&
    	\left.+\frac{m_{\sigma}^{3}}{8\pi M_{D^{*}}^{2}}g_{\Xi_{c}\Xi_{c}\sigma}g_{\sigma}H_{2}\left(i\boldsymbol{\epsilon}_{2} \times \boldsymbol{\epsilon}_{4}^{\dagger}\right)\cdot\boldsymbol{L}
    	+\frac{m_{\sigma}^{3}}{16\pi M_{\Xi_{c}}^{2}}g_{\Xi_{c}\Xi_{c}\sigma}g_{\sigma}H_{2}\left(\boldsymbol{\epsilon}_{2}\cdot\boldsymbol{\epsilon}_{4}^{\dagger}\right)\left(\boldsymbol{\sigma}\cdot\boldsymbol{L}\right)\right],
    \end{align}
    while the vector ($\rho$,$\omega$) exchange potential reads
    \begin{align}\label{eq:24}  	V^{Lab[\Xi_{c}\bar{D}^{*}]}_{vector}\left(\boldsymbol{r},\boldsymbol{\nabla}\right)
    	=&
    	\sum_{\alpha}\mathcal{C}_{\alpha}\left\{
    	\frac{\sqrt{2}m_{v}}{8\pi}\beta g_{V}g_{\Xi_{c}\Xi_{c}V}H_{0}\left(\boldsymbol{\epsilon}_{2}\cdot\boldsymbol{\epsilon}_{4}^{\dagger}\right)
    	-\frac{\sqrt{2}m_{v}^{3}}{12\pi M_{D^{*}}}g_{\Xi_{c}\Xi_{c}V}\lambda g_{V}H_{3}S(\hat{\boldsymbol{r}},\boldsymbol{\epsilon}_{2},\boldsymbol{\epsilon}_{4}^{\dagger})
    	+\frac{\sqrt{2}m_{v}^{3}}{12\pi M_{D^{*}}}g_{\Xi_{c}\Xi_{c}V}\lambda g_{V}H_{1}\left(\boldsymbol{\epsilon}_{2}\cdot\boldsymbol{\epsilon}_{4}^{\dagger}\right)\right.\nonumber\\
    	&
    	\left.+\frac{\sqrt{2}m_{v}^{3}}{4\pi M_{D^{*}}}g_{\Xi_{c}\Xi_{c}V}\lambda g_{V}H_{2}\left[\left(i\boldsymbol{\epsilon}_{2}\times \boldsymbol{\epsilon}_{4}^{\dagger}\right)\cdot\boldsymbol{L}\right]
    	+\frac{\sqrt{2}m_{v}^{3}}{4\pi M_{\Xi_{c}}}g_{\Xi_{c}\Xi_{c}V}\lambda g_{V}H_{2}\left[\left(i\boldsymbol{\epsilon}_{2}\times \boldsymbol{\epsilon}_{4}^{\dagger}\right)\cdot\boldsymbol{L}\right]\right.\nonumber\\
    	&
    	\left.+\frac{\sqrt{2}m_{v}^{3}}{24\pi M_{\Xi_{c}}}g_{\Xi_{c}\Xi_{c}V}\lambda g_{V}H_{3}S(\hat{\boldsymbol{r}}, i\boldsymbol{\epsilon}_{2}\times \boldsymbol{\epsilon}_{4}^{\dagger},\boldsymbol{\sigma})
    	+\frac{\sqrt{2}m_{v}^{3}}{12\pi M_{\Xi_{c}}}g_{\Xi_{c}\Xi_{c}V}\lambda g_{V}H_{1}\left[\left(i\boldsymbol{\epsilon}_{2}\times \boldsymbol{\epsilon}_{4}^{\dagger}\right)\cdot\boldsymbol{\sigma}\right]\right.\nonumber\\
    	&
    	\left.+\frac{\sqrt{2}m_{v}^{3}}{24\pi M_{\Xi_{c}}}f_{\Xi_{c}\Xi_{c}V}\lambda g_{V}H_{3}S(\hat{\boldsymbol{r}}, i\boldsymbol{\epsilon}_{2}\times \boldsymbol{\epsilon}_{4}^{\dagger},\boldsymbol{\sigma})
    	+\frac{\sqrt{2}m_{v}^{3}}{12\pi M_{\Xi_{c}}}f_{\Xi_{c}\Xi_{c}V}\lambda g_{V}H_{1}\left[\left(i\boldsymbol{\epsilon}_{2}\times \boldsymbol{\epsilon}_{4}^{\dagger}\right)\cdot\boldsymbol{\sigma}\right]\right.\nonumber\\
    	&
    	\left.-\frac{\sqrt{2}m_{v}^{3}}{64\pi M_{\Xi_{c}}^{2}}\beta g_{V}g_{\Xi_{c}\Xi_{c}V}H_{1}\left(\boldsymbol{\epsilon}_{2}\cdot\boldsymbol{\epsilon}_{4}^{\dagger}\right)
    	+\frac{\sqrt{2}m_{v}^{3}}{32\pi M_{\Xi_{c}} M_{D^{*}}}\beta g_{V}g_{\Xi_{c}\Xi_{c}V}H_{1}\left(\boldsymbol{\epsilon}_{2}\cdot\boldsymbol{\epsilon}_{4}^{\dagger}\right)
    	-\frac{\sqrt{2}m_{v}^{3}}{32\pi M_{\Xi_{c}}^{2}}\beta g_{V}f_{\Xi_{c}\Xi_{c}V}H_{1}\left(\boldsymbol{\epsilon}_{2}\cdot\boldsymbol{\epsilon}_{4}^{\dagger}\right)\right.\nonumber\\
    	&
    	\left.-\frac{\sqrt{2}m_{v}^{3}}{8\pi M_{\Xi_{c}} M_{D^{*}}}\beta g_{V}g_{\Xi_{c}\Xi_{c}V}H_{2}\left(\boldsymbol{\epsilon}_{2}\cdot\boldsymbol{\epsilon}_{4}^{\dagger}\right)\left(\boldsymbol{r}\cdot \boldsymbol{\nabla}\right)
    	-\frac{\sqrt{2}m_{v}}{8\pi M_{\Xi_{c}} M_{D^{*}}}\beta g_{V}g_{\Xi_{c}\Xi_{c}V}H_{0}\left(\boldsymbol{\epsilon}_{2}\cdot\boldsymbol{\epsilon}_{4}^{\dagger}\right)\boldsymbol{\nabla}^2\right.\nonumber\\
    	&
    	\left.-\frac{\sqrt{2}m_{v}^{3}}{32\pi M_{\Xi_{c}^{2}}}\beta g_{V}g_{\Xi_{c}\Xi_{c}V}H_{2}\left(\boldsymbol{\epsilon}_{2}\cdot\boldsymbol{\epsilon}_{4}^{\dagger}\right)\left(\boldsymbol{\sigma}\cdot \boldsymbol{L}\right)
    	-\frac{\sqrt{2}m_{v}^{3}}{16\pi M_{\Xi_{c}} M_{D^{*}}}\beta g_{V}g_{\Xi_{c}\Xi_{c}V}H_{2}\left(\boldsymbol{\epsilon}_{2}\cdot\boldsymbol{\epsilon}_{4}^{\dagger}\right)\left(\boldsymbol{\sigma}\cdot \boldsymbol{L}\right) \right.\nonumber\\
    	&
    	\left.-\frac{\sqrt{2}m_{v}^{3}}{16\pi M_{\Xi_{c}}^{2}}\beta g_{V}f_{\Xi_{c}\Xi_{c}V}H_{2}\left(\boldsymbol{\epsilon}_{2}\cdot\boldsymbol{\epsilon}_{4}^{\dagger}\right)\left(\boldsymbol{\sigma}\cdot \boldsymbol{L}\right)  
    	-\frac{\sqrt{2}m_{v}^{3}}{16\pi M_{\Xi_{c}}M_{D^{*}}}\beta g_{V}f_{\Xi_{c}\Xi_{c}V}H_{2}\left(\boldsymbol{\epsilon}_{2}\cdot\boldsymbol{\epsilon}_{4}^{\dagger}\right)\left(\boldsymbol{\sigma}\cdot \boldsymbol{L}\right)\right.\nonumber\\
    	&
    	\left.-\frac{\sqrt{2}m_{v}^{3}}{48\pi M_{D^{*}}^{2}}\beta g_{V}g_{\Xi_{c}\Xi_{c}V}H_{3}S(\hat{\boldsymbol{r}}, \boldsymbol{\epsilon}_{2}, \boldsymbol{\epsilon}_{4}^{\dagger})
    	+\frac{\sqrt{2}m_{v}^{3}}{48\pi M_{D^{*}}^{2}}\beta g_{V}g_{\Xi_{c}\Xi_{c}V}H_{1}\left(\boldsymbol{\epsilon}_2 \cdot \boldsymbol{\epsilon}_{4}^{\dagger}\right)\right.\nonumber\\
    	&
    	\left.+\frac{\sqrt{2}m_{v}^{3}}{16\pi M_{D^{*}}^{2}}\beta g_{V}g_{\Xi_{c}\Xi_{c}V}H_{2}\left[\left(i\boldsymbol{\epsilon}_{2} \times\boldsymbol{\epsilon}_{4}^{\dagger}\right)\cdot\boldsymbol{L}\right]\right\},	    	
    \end{align}
    where $\alpha=\rho,\omega$.
    \par
    $\mathcal{C}_{\sigma}$ and $\mathcal{C}_{\alpha}$ in the Eqs. (\ref{eq:21})\--(\ref{eq:24}) are isospin factors derived from the isospin wave functions. 
    The definition of the tensor operator $S(\hat{\boldsymbol{r}},\boldsymbol{a},\boldsymbol{b})$ and the expressions of $H_{0}$, $H_{1}$, $H_{2}$ and $H_{3}$ are given in the Appendix.
    \par 
    $G$-parity rule states that the one-boson-exchange amplitude (equivalent to the effective potentials) of the $AB$ to $AB$ scattering process is linked with that of the $A\bar{B}$ to $A\bar{B}$ one-boson-exchange scattering process by a factor $(-)^{I_{G}}$, where $(-)^{I_{G}}$ is the $G$ parity of the exchanged light meson \cite{ref43,ref49}. Consequently, the effective potentials of the open charm systems read
    \begin{equation}
    	\begin{aligned}
    		V^{Lab[\Xi_{c}D^{(*)}]}_{scalar}=(-)^{I_{G}[\sigma]}V^{Lab[\Xi_{c}\bar{D}^{(*)}]}_{scalar},\\
    		V^{Lab[\Xi_{c}D^{(*)}]}_{vector}=(-)^{I_{G}[\alpha]}V^{Lab[\Xi_{c}\bar{D}^{(*)}]}_{vector},\\
    		V^{Lab[\Lambda_{c}D^{(*)}]}_{scalar}=(-)^{I_{G}[\sigma]}V^{Lab[\Lambda_{c}\bar{D}^{(*)}]}_{scalar},\\    		V^{Lab[\Lambda_{c}D^{(*)}]}_{vector}=(-)^{I_{G}[\alpha]}V^{Lab[\Lambda_{c}\bar{D}^{(*)}]}_{vector}.
    	\end{aligned}
    \end{equation}  
    \par
    For practical calculations, we absorb the $(-)^{I_{G}}$ factors of the exchanged mesons into a redefined set of isospin factors for the open-charm systems, which ensures the isospin-independent parts of the potentials remain identical. The isospin factors used in this work are exhibited in TABLE \ref{tab:4}.
    \par 
    When solving the coupled channel Schr\"{o}dinger equation, the structures related to angular momentum will be integrated into the matrix elements, where element corresponds to a set of initial and final partial waves. The matrix elements are given in the Appendix.
     \renewcommand{\arraystretch}{1.5}
    \begin{table*}[!htb]
    	\scriptsize
    	\begin{center}
    		\caption{\label{tab:4}  The isospin factors for different channels}
    		
    		\begin{tabular}{ c|ccc}\toprule[1pt]
    			\multicolumn{1}{c|}{Channels}  &
    			\multicolumn{1}{c}{$C_\rho$} &
    			\multicolumn{1}{c}{$C_\omega$} &
    			\multicolumn{1}{c}{$C_\sigma$} \\
    			\midrule[1pt]
    			$[\Xi_{c}\bar{D}^{(*)}]_{I=1}\longleftrightarrow [\Xi_{c}\bar{D}^{(*)}]_{I=1}$ & $\frac{1}{2}$ & $\frac{1}{2}$ & $1$  \\				
    			$[\Xi_{c}\bar{D}^{(*)}]_{I=0}\longleftrightarrow [\Xi_{c}\bar{D}^{(*)}]_{I=0}$ & $-\frac{3}{2}$ & $\frac{1}{2}$ & $1$  \\		
    			$[\Xi_{c}D^{(*)}]_{I=1}\longleftrightarrow [\Xi_{c}D^{(*)}]_{I=1}$ & $\frac{1}{2}$ & $-\frac{1}{2}$ & $1$  \\				
    			$[\Xi_{c}D^{(*)}]_{I=0}\longleftrightarrow [\Xi_{c}D^{(*)}]_{I=0}$ & $-\frac{3}{2}$ & $-\frac{1}{2}$ & $1$  \\	
    			$[\Lambda_{c}^{+}\bar{D}^{(*)}]_{I=\frac{1}{2}}\longleftrightarrow [\Lambda_{c}^{+}\bar{D}^{(*)}]_{I=\frac{1}{2}}$ & $0$ & $\frac{1}{2}$ & $1$  \\
    			$[\Lambda_{c}^{+}D^{(*)}]_{I=\frac{1}{2}}\longleftrightarrow [\Lambda_{c}^{+}\bar{D}^{(*)}]_{I=\frac{1}{2}}$ & $0$ & $-\frac{1}{2}$ & $1$  \\
    			\bottomrule[1pt]
    		\end{tabular}
    	\end{center}
    \end{table*}

	\section{Numerical results} \label{sec3}
	 With the effective potentials derived in the previous chapter, we obtain the bound state solutions by solving the coupled channel Schr\"{o}dinger equation numerically. The computed values of characteristic quantities such as the binding energy and the root-mean-square (RMS) radius are listed in the following tables. In the loosely bound picture, a plausible hadronic molecular candidate is expected to have a binding energy of no more than a few tens of MeV and a root-mean-square radius of approximately 1 fm.
	\par 
	 Besides, the cutoff value of the form factor affects the judgment of the hadronic molecular states. Unfortunately, due to the non-perturbative nature of QCD in the low energy region and the lack of relevant experimental data, the cutoff value still cannot be precisely determined. In practice, researchers typically fix the range of cutoff values based on the experience from the studies of the deuteron. Since the typical value of the cutoff parameter for the deuteron lies in the range of 1.2\--1.5 GeV, it is generally believed that the cutoff value for other hadronic molecules should not deviate significantly from 1 GeV. In this study, we therefore adopt a cutoff value range of 0.8\--2.0 GeV.
	 
	\subsection{The mass spectra of $\Xi_{c}\bar{D}^{(*)}$ systems} \label{subsecD}
    The constraint of angular momentum conservation requires that the $0(\frac{1}{2}^{-})$ channel of the $\Xi_{c}\bar{D}$ system in its ground state must be pure $S$ wave, while the $0(\frac{1}{2}^{-})$ and $0(\frac{3}{2}^{-})$ channels of the $\Xi_{c}\bar{D}^{*}$ system gain a few $D$ wave components in their ground states with recoil corrections. 
    \par 
    For the isospin triplet channels, the effective potentials always provide a repulsive interaction, which leads to no bound state. For the isospin singlet channels, we collect the numerical results of binding energy and the root-mean-square radius with different cutoff value in TABLE \ref{tab:5}. In order to exhibit the influence of recoil corrections to certain channels, the results with and without the momentum-related terms are given separately.  
	\par
\begin{figure}[!htbp]
		\centering
		\begin{subfigure}{0.48\textwidth}
			\centering
			\includegraphics[scale=0.4]{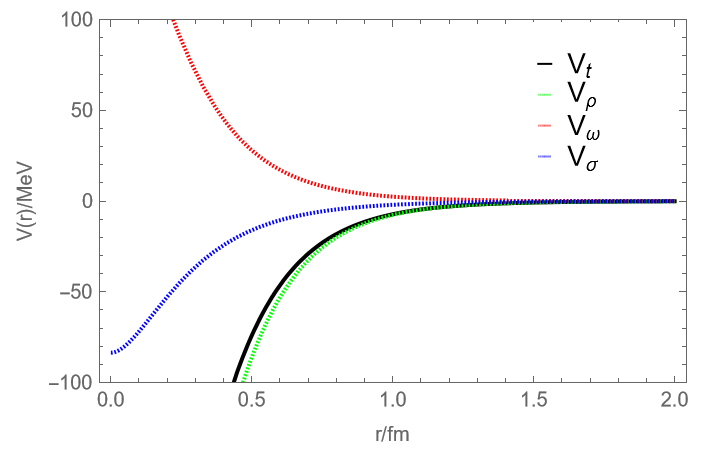}
			\caption{}
			\label{subfig:1}
		\end{subfigure}
		\hfill
		\begin{subfigure}{0.48\textwidth}
			\centering
			\includegraphics[scale=0.4]{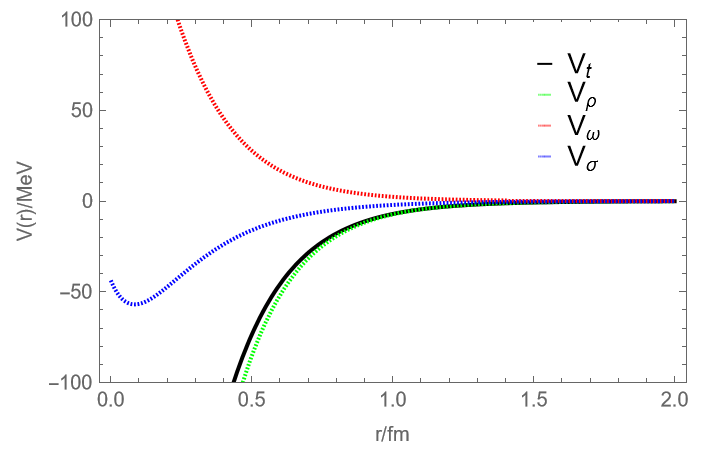}
			\caption{}
			\label{subfig:2}
		\end{subfigure}
		\captionsetup{format=plain, justification=raggedright, singlelinecheck=false}
		\caption{The effective potentials for the $\Xi_{c}\bar{D}$ system in the $0(\frac{1}{2}^{-})$channel. Panel (a): contributions from $\sigma$, $\rho$, and $\omega$ exchange without recoil corrections. Panel (b): same with recoil corrections. The cutoff value is $\Lambda=1.75$ GeV.}
		\label{fig:1}
	    \end{figure}

	The effective potentials for the $0(\frac{1}{2}^{-})$ channel of the $\Xi_{c}\bar{D}$ system are presented in FIG. \ref{fig:1}. Panel~(a) displays the $S$ wave potentials without recoil corrections, while Panel~(b) displays those with recoil corrections, To clarify the contribution of each meson and to offer a comprehensive perspective, we show both the individual meson contributions and the total potential in each panel.
	\par
	As we can see in TABLE \ref{tab:5} and FIG. \ref{fig:1}, a bound state solution appears when the cutoff parameter $\Lambda=1.55$ GeV. The $\rho$ and $\sigma$ exchanges provide the attractive interaction. For a given $\Lambda$, the inclusion of recoil corrections slightly reduces the binding energy and increases the RMS radius, which indicates that such corrections, though inconspicuous, make the bound state weaker and more extended.
	\par
\begin{figure}[!htbp]
	    	\centering
	    	\begin{subfigure}{0.32\textwidth}
	    		\centering
	    		\includegraphics[scale=0.4]{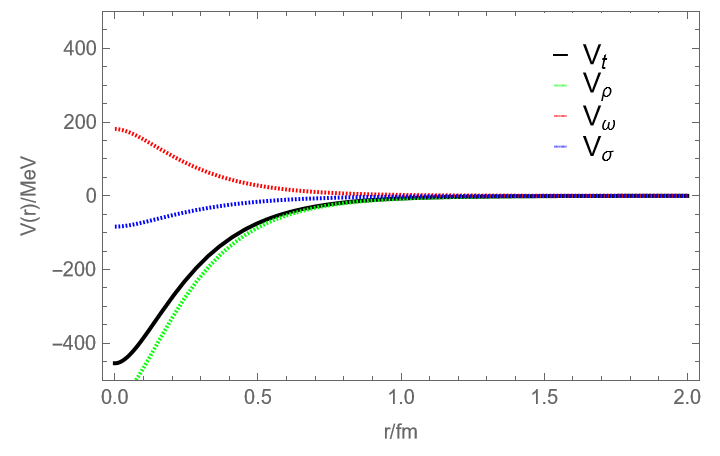}
	    		\label{subfig:3}
	    		\caption{}
	    	\end{subfigure}
	    	\hfill
	    	\begin{subfigure}{0.32\textwidth}
	    		\centering
	    		\includegraphics[scale=0.4]{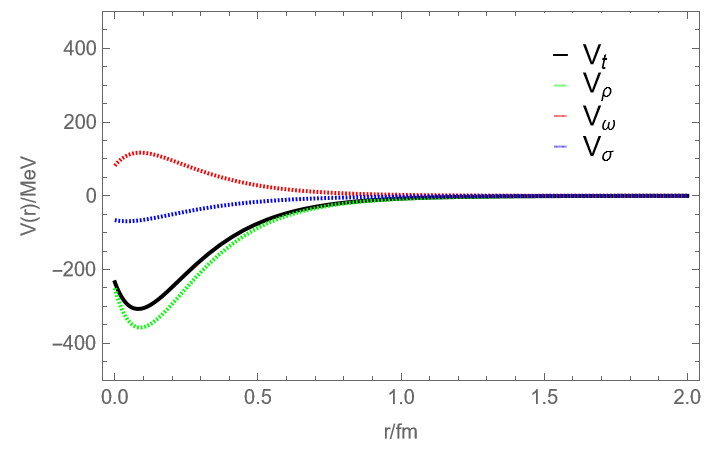}
	    		\label{subfig:4}
	    		\caption{}
	    	\end{subfigure}
	    	\hfill
	    	\begin{subfigure}{0.32\textwidth}
	    		\centering
	    		\includegraphics[scale=0.4]{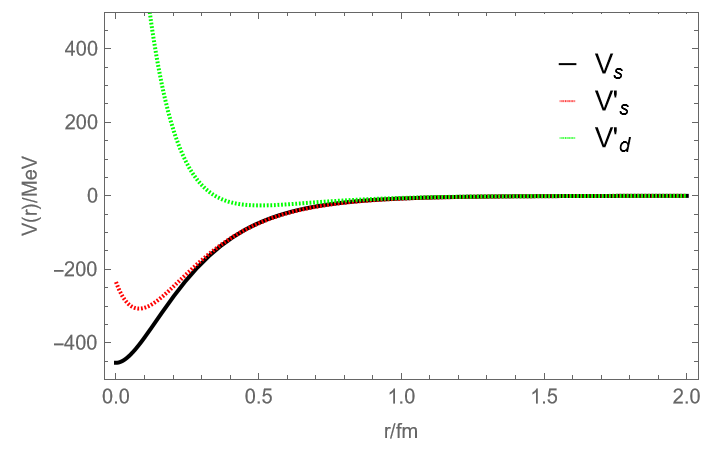}
	    		\label{subfig:5}
	    		\caption{}
	    	\end{subfigure}
	    	\hfill
	    	\begin{subfigure}{0.32\textwidth}
	    		\centering
	    		\includegraphics[scale=0.4]{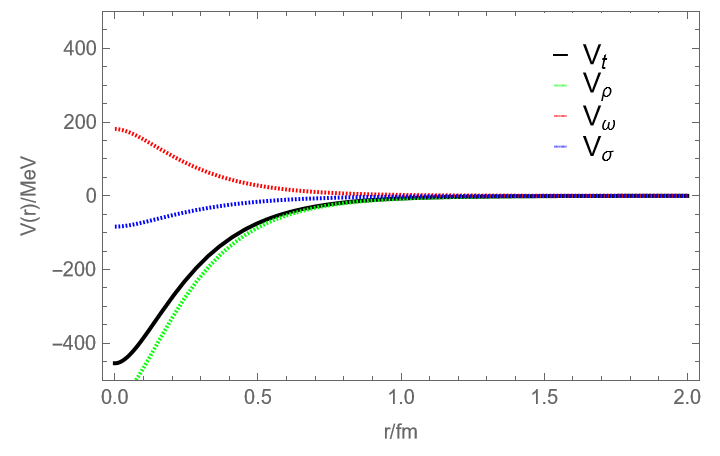}
	    		\label{subfig:6}
	    		\caption{}
	    	\end{subfigure}
	    	\hfill
	    	\begin{subfigure}{0.32\textwidth}
	    		\centering
	    		\includegraphics[scale=0.4]{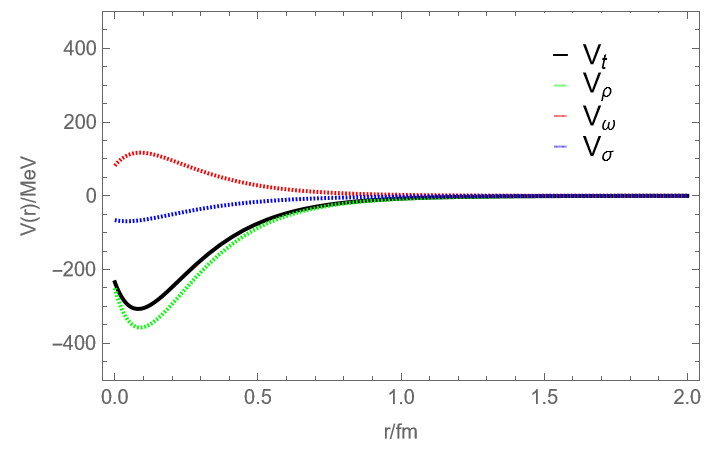}
	    		\label{subfig:7}
	    		\caption{}
	    	\end{subfigure}
	    	\hfill
	    	\begin{subfigure}{0.32\textwidth}
	    		\centering
	    		\includegraphics[scale=0.4]{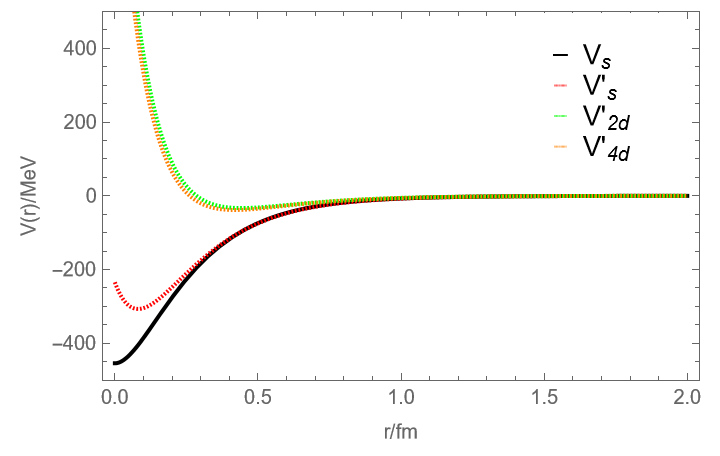}
	    		\label{subfig:8}
	    		\caption{}
	    	\end{subfigure}
	    	\captionsetup{format=plain, justification=raggedright, singlelinecheck=false}
	    	\caption{The effective potentials for the $\Xi_{c}\bar{D}^{*}$ system in the $0(\frac{1}{2}^{-})$ (upper row) and $0(\frac{3}{2}^{-})$  (lower row) channels. Panels (a,d): contributions from $\sigma$, $\rho$, and $\omega$ exchange without recoil corrections. Panels (b,e): same with recoil corrections. Panels (c,f): comparison of the total potentials in the ground state with and without recoil corrections. The cutoff value is $\Lambda=1.75$ GeV.}
	    	\label{fig:2}
	    \end{figure}

 
	The effective potentials for channels of the $\Xi_{c}\bar{D}^{*}$ system are presented in FIG. \ref{fig:2}, where the upper and lower rows correspond to the $0(\frac{1}{2}^{-})$ and $0(\frac{3}{2}^{-})$ channels, respectively. Within each row, the first two panels [Panels (a) and (b) above; (d) and (e) below] display the same physical objects as Panels (a) and (b) in FIG. \ref{fig:1}. Since recoil corrections are noticeable in these channels, the last panel [Panel (c) above; (f) below] compare the overall effective potentials with and without the recoil corrections.
	\par
	As evident from TABLE \ref{tab:5} and FIG. \ref{fig:2}, compare with the case of the $\Xi_{c}\bar{D}$ system, recoil corrections play a more significant role in the $\Xi_{c}\bar{D}^{*}$ system, markedly altering the spectrum of each channel. For instance, in the $0(\frac{1}{2}^{-})$ channel with $\Lambda=1.75$ GeV, a bound state with binding energy $5.34$ MeV emerges without recoil corrections. However, if we take such corrections into consideration, the binding energy decreases to $3.08$ MeV (just over half). Compare the Panel (a) and the Panel (b) in FIG. \ref{fig:2}, recoil corrections shallow the $\rho$ exchange potential in the $S$ wave, especially in the short-distance region. Now that the $\rho$ exchange potential is the major donor for the attractive interaction, similar changes in the total $S$ wave potential is exhibited in the Panel (c). Another noteworthy phenomenon is that even if the recoil corrections introduce $D$ wave components into the ground state, the proportion of these components are so small that the split of the $0(\frac{1}{2}^{-})$ and $0(\frac{3}{2}^{-})$ channels can hardly be observed.

		\begin{table}[htbp]
			\scriptsize
			\begin{center}
				\caption{\label{tab:5} The numerical results of the  $\Xi_{c}\bar{D}$ and $\Xi_{c}\bar{D}^{*}$ systems in different channels}
				\begin{tabular}{l ccc|cccccc}\toprule[1pt]
					\multicolumn{10}{c}{$\Xi_{c}\bar{D}$ [$I(J^P)=0(\frac{1}{2}^-)$] }\\
					\midrule[1pt]
					\multicolumn{4}{c|}{Without recoil corrections} & \multicolumn{6}{c}{With recoil corrections}\\
					\midrule[1pt]
					$\Lambda$(GeV) & B.E.(MeV) & RMS(fm)  & $^2S_{\frac{1}{2}} $($\%$)  & $\Lambda$(GeV) & B.E.(MeV) & RMS(fm)  & $^2S_{\frac{1}{2}} $($\%$) & &  \\  
					1.55 & 0.28 & 5.23 & 100 & 1.55 & 0.21 & 5.57 & 100 & &  \\
					1.60 & 0.84 & 3.55 & 100 & 1.60 & 0.71 & 3.79 & 100 & & \\
					1.65 & 1.65 & 2.64 & 100 & 1.65 & 1.46 & 2.79 & 100 & & \\
					1.70 & 2.70 & 2.12 & 100 & 1.70 & 2.42 & 2.23 & 100 & & \\
					1.75 & 3.96 & 1.80 & 100 & 1.75 & 3.58 & 1.88 & 100 & & \\
					\midrule[1pt]
					\multicolumn{10}{c}{$\Xi_{c}\bar{D}^{*}$ [$I(J^P)=0(\frac{1}{2}^-)$] }\\
					\midrule[1pt]
					\multicolumn{4}{c|}{Without recoil corrections} & \multicolumn{6}{c}{With recoil corrections}\\
					\midrule[1pt]
					$\Lambda$(GeV) & B.E.(MeV) & RMS(fm)  & $^2S_{\frac{1}{2}} $($\%$)  & $\Lambda$(GeV) & B.E.(MeV) & RMS(fm)  & $^2S_{\frac{1}{2}} $($\%$) & $^4D_{\frac{1}{2}} $($\%$) & \\  
					1.55 & 0.68 & 3.81 & 100 & 1.55 & 0.22 & 5.50 & 99.99 & 0.01 &\\
					1.60 & 1.47 & 2.73 & 100 & 1.60 & 0.67 & 3.85 & 99.99 & 0.01 &\\
					1.65 & 2.53 & 2.15 & 100 & 1.65 & 1.31 & 2.89 & 99.98 & 0.02 &\\
					1.70 & 3.83 & 1.80 & 100 & 1.70 & 2.12 & 2.34 & 99.98 & 0.02 &\\
					1.75 & 5.34 & 1.56 & 100 & 1.75 & 3.08 & 1.99 & 99.97 & 0.03 &\\
					\midrule[1pt]
					\multicolumn{10}{c}{$\Xi_{c}\bar{D}^{*}$ [$I(J^P)=0(\frac{3}{2}^-)$] }\\
					\midrule[1pt]
					\multicolumn{4}{c|}{Without recoil corrections} & \multicolumn{6}{c}{With recoil corrections}\\
					\midrule[1pt]
					$\Lambda$(GeV) & B.E.(MeV) & RMS(fm)  & $^4S_{\frac{3}{2}} $($\%$)  & $\Lambda$(GeV) & B.E.(MeV) & RMS(fm)  & $^4S_{\frac{3}{2}} $($\%$) & $^2D_{\frac{3}{2}} $($\%$) & $^4D_{\frac{3}{2}} $($\%$) \\  
					1.55 & 0.68 & 3.81 & 100 & 1.55 & 0.22 & 5.50 & 99.99 & 0.00 & 0.01 \\
					1.60 & 1.47 & 2.73 & 100 & 1.60 & 0.67 & 3.85 & 99.99 & 0.00 & 0.01 \\
					1.65 & 2.53 & 2.15 & 100 & 1.65 & 1.31 & 2.89 & 99.98 & 0.01 & 0.01 \\
					1.70 & 3.83 & 1.80 & 100 & 1.70 & 2.12 & 2.34 & 99.98 & 0.01 & 0.01 \\
					1.75 & 5.34 & 1.56 & 100 & 1.75 & 3.07 & 1.99 & 99.97 & 0.01 & 0.02 \\
					\bottomrule[1pt]								
				\end{tabular}
			\end{center}
		\end{table} 
		\clearpage

    	\clearpage

	\subsection{The mass spectra of $\Xi_{c}D^{(*)}$ systems} \label{subsecE}
	Similar with the $\Xi_{c}D^{(*)}$ systems, in the $0(\frac{1}{2}^{-})$ channel of the $\Xi_{c}D$ system, the ground state only contains $S$ wave; in the $0(\frac{1}{2}^{-})$ and $0(\frac{3}{2}^{-})$ channels of the $\Xi_{c}D^{*}$ system, recoil corrections introduce the $D$ wave components into the ground states.
	\par 
	For the isospin triplet channels, the effective potentials derived from $\rho$ and $\omega$ exchanges almost cancel out, as they possess opposite signs and comparable strengths. Since there are no pseudoscalar exchanges, the total $S$ wave effective potential is predominantly governed by $\sigma$ exchange. Though attractive, this contribution is insufficient to support the formation of a bound state. For the isospin singlet channels, we collect the numerical results with and without the recoil corrections in TABLE \ref{tab:6}.
	\par
\begin{figure}[!htbp]
		\centering
		\begin{subfigure}{0.48\textwidth}
			\centering
			\includegraphics[scale=0.4]{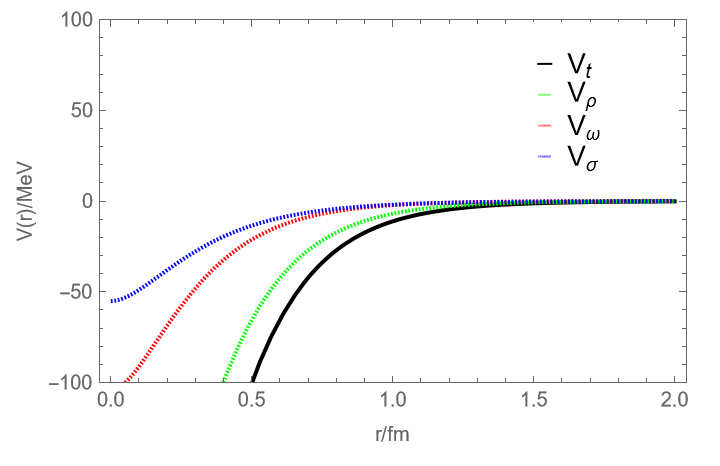}
			\label{subfig:9}
			\caption{}
		\end{subfigure}
		\hfill
		\begin{subfigure}{0.48\textwidth}
			\centering
			\includegraphics[scale=0.4]{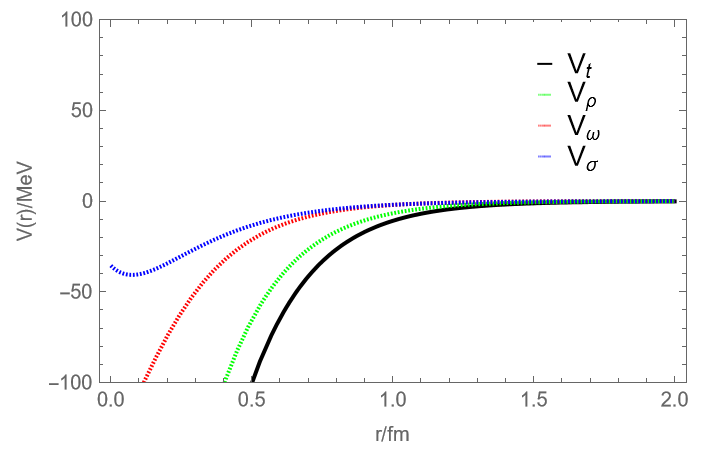}
			\label{subfig:10}
			\caption{}
		\end{subfigure}
		\captionsetup{format=plain, justification=raggedright, singlelinecheck=false}
		\caption{The effective potentials for the $\Xi_{c}D$ system in the $0(\frac{1}{2}^{-})$channel. Panel (a): contributions from $\sigma$, $\rho$, and $\omega$ exchange without recoil corrections. Panel (b): same with recoil corrections. The cutoff value is $\Lambda=1.45$ GeV.}
		\label{fig:3}
	\end{figure}

	The effective potentials for the $0(\frac{1}{2}^{-})$ channel of the $\Xi_{c}D$ system are presented in FIG. \ref{fig:3}. Panels (a) and (b) are the same as in FIG. \ref{fig:1}
	\par
	As we can see in TABLE \ref{tab:6} and FIG. \ref{fig:3}, each of the three exchanged light mesons offers an attractive potential. Thus, it is easier for this channel to form a stable bound state compared with the same channel in its hidden charm partner. Without the recoil corrections, a bound state solution with binding energy 1.17\--17.3 MeV , and with corresponding RMS radius 3.11\--1.01 fm appears when the cutoff value is chosen between 1.25 and 1.45 GeV. The recoil corrections in this channel are unfavorable to the formation of bound states. For the same range of $\Lambda$, the binding energy reduces to 1.05\--16.2 MeV, while the RMS radius increases to 3.26\--1.04 fm.
	\par
\begin{figure}[!htbp]
		\centering
		\begin{subfigure}{0.32\textwidth}
			\centering
			\includegraphics[scale=0.4]{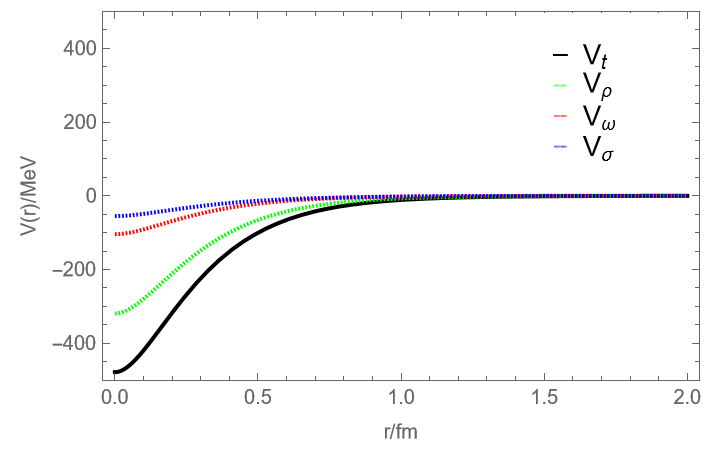}
			\label{subfig:11}
			\caption{}
		\end{subfigure}
		\hfill
		\begin{subfigure}{0.32\textwidth}
			\centering
			\includegraphics[scale=0.4]{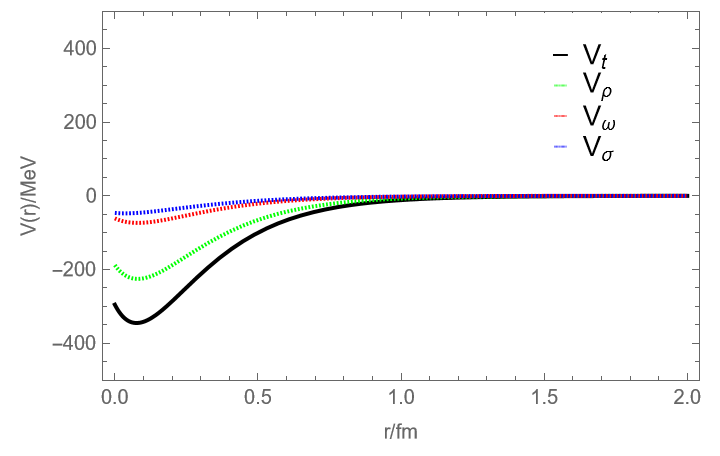}
			\label{subfig:12}
			\caption{}
		\end{subfigure}
		\hfill
		\begin{subfigure}{0.32\textwidth}
			\centering
			\includegraphics[scale=0.4]{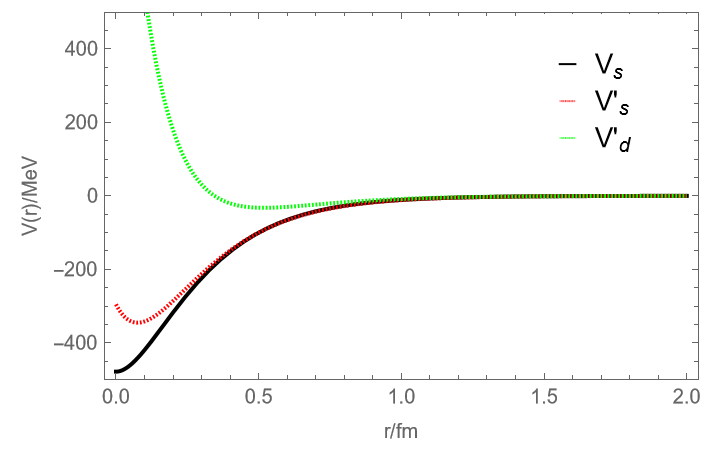}
			\label{subfig:13}
			\caption{}
		\end{subfigure}
		\hfill
		\begin{subfigure}{0.32\textwidth}
			\centering
			\includegraphics[scale=0.4]{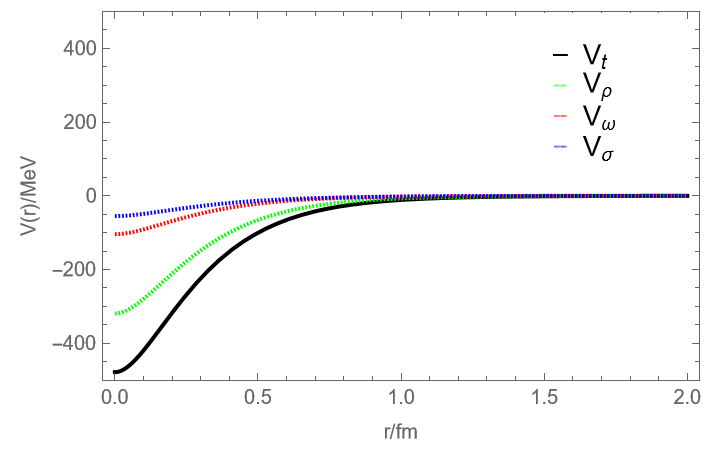}
			\label{subfig:14}
			\caption{}
		\end{subfigure}
		\hfill
		\begin{subfigure}{0.32\textwidth}
			\centering
			\includegraphics[scale=0.4]{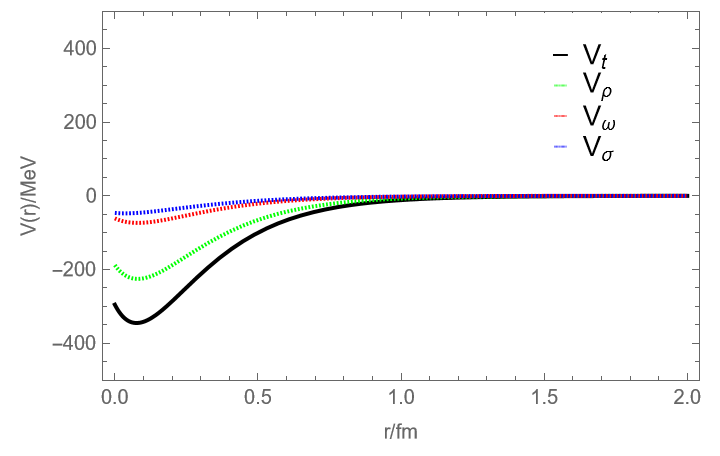}
			\label{subfig:15}
			\caption{}
		\end{subfigure}
		\hfill
		\begin{subfigure}{0.32\textwidth}
			\centering
			\includegraphics[scale=0.4]{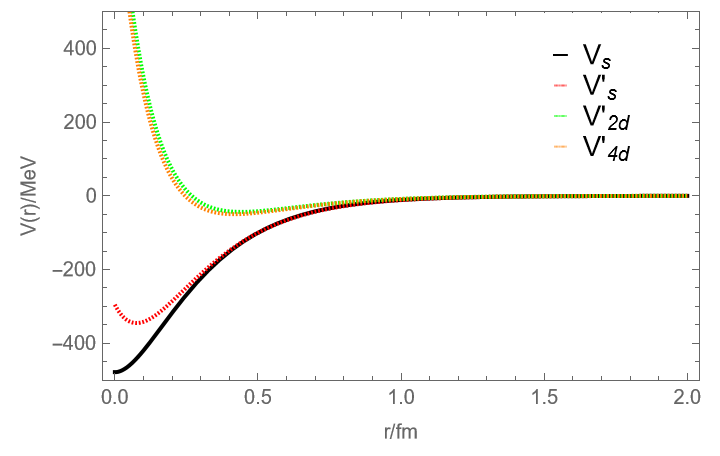}
			\label{subfig:16}
			\caption{}
		\end{subfigure}
		\captionsetup{format=plain, justification=raggedright, singlelinecheck=false}
		\caption{The effective potentials for the $\Xi_{c}D^{*}$ system in the $0(\frac{1}{2}^{-})$ (upper row) and $0(\frac{3}{2}^{-})$  (lower row) channels. Panels (a,d): contributions from $\sigma$, $\rho$, and $\omega$ exchange without recoil corrections. Panels (b,e): same with recoil corrections. Panels (c,f): comparison of the total potentials in the ground state with and without recoil corrections. The cutoff value is $\Lambda=1.45$ GeV.}
		\label{fig:4}
	\end{figure}

 
	The effective potentials for channels of the $\Xi_{c}D^{*}$ system are presented in FIG. \ref{fig:4}, where the upper and lower rows correspond to the $0(\frac{1}{2}^{-})$ and $0(\frac{3}{2}^{-})$ channels, respectively. Panels (a) to (f) are the same as in FIG. \ref{fig:2}
	\par
	Comparable to the channels of the $\Xi_{c}\bar{D}^{*}$ system, recoil corrections show striking influence on the spectrum of each channel of the $\Xi_{c}D^{*}$ system. It is exhibited in TABLE \ref{tab:6} and FIG. \ref{fig:4} that such corrections are also unfavorable for the formation of a bound state because the $\rho$ and $\omega$ exchange potentials become shallower. In the $0(\frac{1}{2}^{-})$ channel, when the cutoff parameter is fixed as $\Lambda=1.45$ GeV, a bound state with binding energy $19.9$ MeV emerges without recoil corrections, which substantially decreases to $14.0$ MeV with recoil corrections. 
	 \begin{table*}[htbp]
		\scriptsize
		\begin{center}
			\caption{\label{tab:6} The numerical results of the  $\Xi_{c}D$ and $\Xi_{c}D^{*}$ systems in different channels}
			\begin{tabular}{l ccc|cccccc}\toprule[1pt]
				\multicolumn{10}{c}{$\Xi_{c}D$ [$I(J^P)=0(\frac{1}{2}^-)$] }\\
				\midrule[1pt]
				\multicolumn{4}{c|}{Without recoil corrections} & \multicolumn{6}{c}{With recoil corrections}\\
				\midrule[1pt]
				$\Lambda$(GeV) & B.E.(MeV) & RMS(fm)  & $^2S_{\frac{1}{2}} $($\%$)  & $\Lambda$(GeV) & B.E.(MeV) & RMS(fm)  & $^2S_{\frac{1}{2}} $($\%$) & &  \\  
				1.25 & 1.17 & 3.11 & 100 & 1.25 & 1.05 & 3.26 & 100 & & \\
				1.30 & 3.58 & 1.91 & 100 & 1.30 & 3.30 & 1.98 & 100 & & \\
				1.35 & 7.14 & 1.44 & 100 & 1.35 & 6.66 & 1.48 & 100 & & \\
				1.40 & 11.7 & 1.18 & 100 & 1.40 & 11.0 & 1.21 & 100 & & \\
				1.45 & 17.3 & 1.01 & 100 & 1.45 & 16.2 & 1.04 & 100 & & \\
				\midrule[1pt]
				\multicolumn{10}{c}{$\Xi_{c}D^{*}$ [$I(J^P)=0(\frac{1}{2}^-)$] }\\
				\midrule[1pt]
				\multicolumn{4}{c|}{Without recoil corrections} & \multicolumn{6}{c}{With recoil corrections}\\
				\midrule[1pt]
				$\Lambda$(GeV) & B.E.(MeV) & RMS(fm)  & $^2S_{\frac{1}{2}} $($\%$)  & $\Lambda$(GeV) & B.E.(MeV) & RMS(fm)  & $^2S_{\frac{1}{2}} $($\%$) & $^4D_{\frac{1}{2}} $($\%$) & \\  
				1.25 & 1.82 & 2.52 & 100 & 1.25 & 1.03 & 3.23 & 99.99 & 0.01 &\\
				1.30 & 4.70 & 1.68 & 100 & 1.30 & 3.05 & 2.03 & 99.98 & 0.02 &\\
				1.35 & 8.76 & 1.30 & 100 & 1.35 & 5.97 & 1.54 & 99.96 & 0.04 &\\
				1.40 & 13.9 & 1.09 & 100 & 1.40 & 9.66 & 1.27 & 99.95 & 0.05 &\\
				1.45 & 19.9 & 0.95 & 100 & 1.45 & 14.0 & 1.10 & 99.93 & 0.07 &\\
				\midrule[1pt]
				\multicolumn{10}{c}{$\Xi_{c}D^{*}$ [$I(J^P)=0(\frac{3}{2}^-)$] }\\
				\midrule[1pt]
				\multicolumn{4}{c|}{Without recoil corrections} & \multicolumn{6}{c}{With recoil corrections}\\
				\midrule[1pt]
				$\Lambda$(GeV) & B.E.(MeV) & RMS(fm)  & $^4S_{\frac{3}{2}} $($\%$)  & $\Lambda$(GeV) & B.E.(MeV) & RMS(fm)  & $^4S_{\frac{3}{2}} $($\%$) & $^2D_{\frac{3}{2}} $($\%$) & $^4D_{\frac{3}{2}} $($\%$) \\  
				1.25 & 1.82 & 2.52 & 100 & 1.25 & 1.03 & 3.23 & 99.99 & 0.00 & 0.01 \\
				1.30 & 4.70 & 1.68 & 100 & 1.30 & 3.05 & 2.03 & 99.98 & 0.01 & 0.01 \\
				1.35 & 8.76 & 1.30 & 100 & 1.35 & 5.97 & 1.54 & 99.96 & 0.02 & 0.02 \\
				1.40 & 13.9 & 1.09 & 100 & 1.40 & 9.66 & 1.27 & 99.95 & 0.02 & 0.03 \\
				1.45 & 19.9 & 0.95 & 100 & 1.45 & 14.0 & 1.10 & 99.93 & 0.03 & 0.04 \\
				\bottomrule[1pt]				
			\end{tabular}
		\end{center}
	\end{table*} 
	\clearpage
	
	\clearpage
	
	\subsection{The mass spectra of $\Lambda_{c}\bar{D}^{(*)}$ and $\Lambda_{c}D^{(*)}$ systems} \label{subsecF}
	When the cutoff value is constrained under the 2.00 GeV, we fail to find a bound state solution in any channels of the  $\Lambda_{c}\bar{D}^{(*)}$ or $\Lambda_{c}D^{(*)}$ system, whether recoil corrections are considered or not.
	
	\subsection{The mass spectra of $\Xi_{b}B^{(*)}$ and $\Xi_{b}\bar{B}^{(*)}$ systems} \label{subsecG}
	The dynamics of binding in $\Xi_{b}B^{(*)}$ and $\Xi_{b}\bar{B}^{(*)}$ systems is qualitatively analogous to that of their charmed counterparts. Therefore, we present relevant results in the Appendix. Due to their larger masses, the bottom-flavor systems are more favorable for the formation of hadronic molecules, while the influence of recoil corrections is undermined.
	\subsection{The mass spectra of $\Lambda_{b}B^{(*)}$ and $\Lambda_{b}\bar{B}^{(*)}$ systems} \label{subsecH}
	None of the channels of the $\Lambda_{b}B^{(*)}$ systems owns a bound state solution. However, unlike the $\Lambda_{c}D$ and $\Lambda_{c}D^{*}$ systems, there exists bound state solutions in the channels of the open bottom systems. We collect the numerical results with and without the recoil corrections in TABLE \ref{tab:7}.
	\par
\begin{figure}[!htbp]
    	\centering
    	\begin{subfigure}{0.48\textwidth}
    		\centering
    		\includegraphics[scale=0.4]{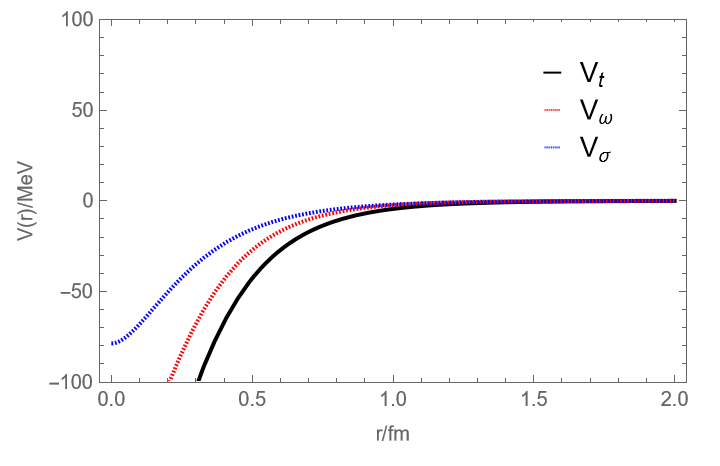}
    		\caption{}
    		\label{subfig:17}
    	\end{subfigure}
    	\hfill
    	\begin{subfigure}{0.48\textwidth}
    		\centering
    		\includegraphics[scale=0.4]{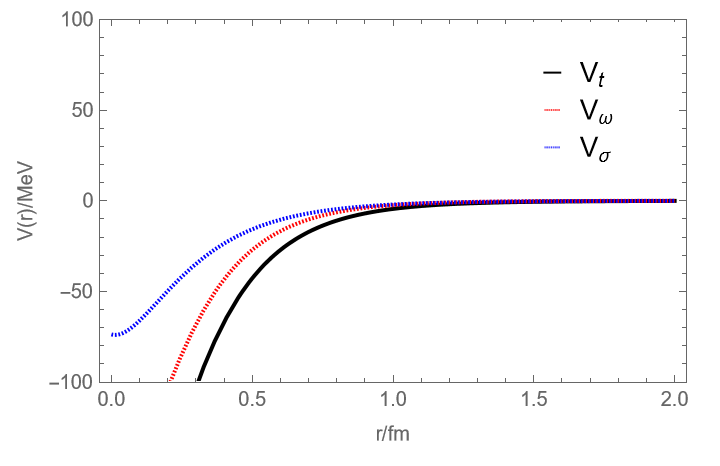}
    		\caption{}
    		\label{subfig:18}
    	\end{subfigure}
    	\captionsetup{format=plain, justification=raggedright, singlelinecheck=false}
    	\caption{The effective potentials for the $\Lambda_{b}\bar{B}$ system in the $\frac{1}{2}(\frac{1}{2}^{-})$channel. Panel (a): contributions from $\sigma$ and $\omega$ exchange without recoil corrections. Panel (b): same with recoil corrections. The cutoff value is $\Lambda=1.70$ GeV.}
    	\label{fig:5}
    \end{figure}
\begin{figure}[!htbp]
    	\centering
    	\begin{subfigure}{0.32\textwidth}
    		\centering
    		\includegraphics[scale=0.4]{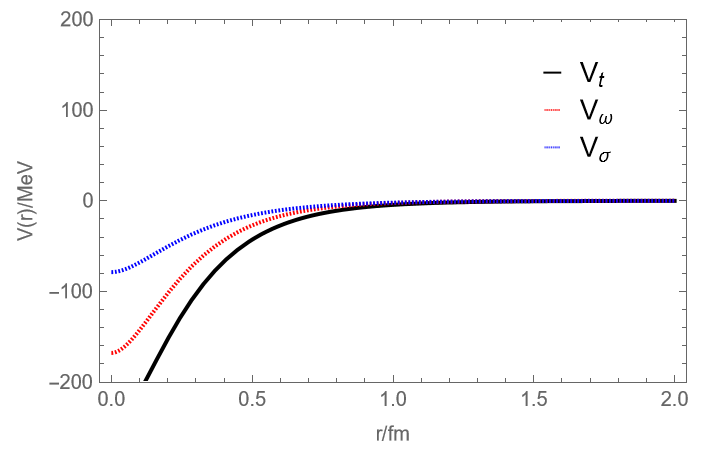}
    		\label{subfig:19}
    		\caption{}
    	\end{subfigure}
    	\hfill
    	\begin{subfigure}{0.32\textwidth}
    		\centering
    		\includegraphics[scale=0.4]{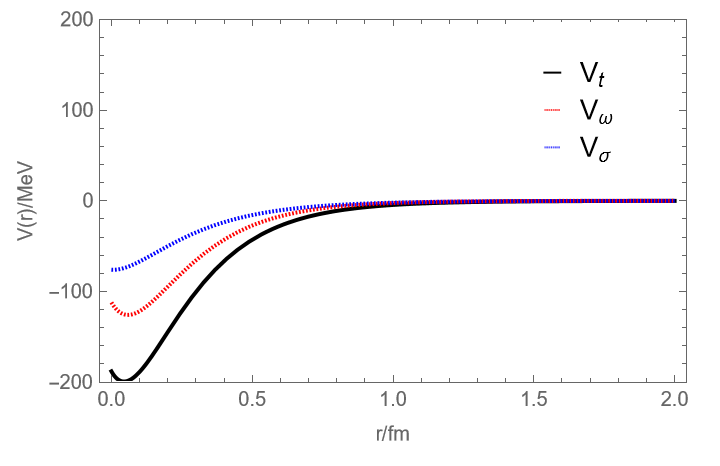}
    		\label{subfig:20}
    		\caption{}
    	\end{subfigure}
    	\hfill
    	\begin{subfigure}{0.32\textwidth}
    		\centering
    		\includegraphics[scale=0.4]{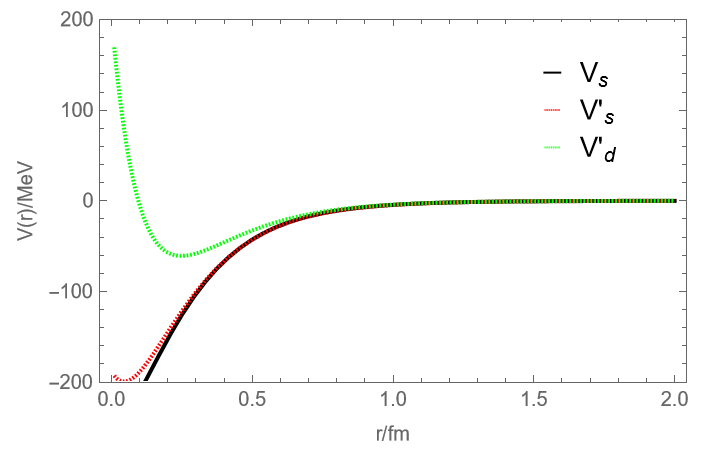}
    		\label{subfig:21}
    		\caption{}
    	\end{subfigure}
    	\hfill
    	\begin{subfigure}{0.32\textwidth}
    		\centering
    		\includegraphics[scale=0.4]{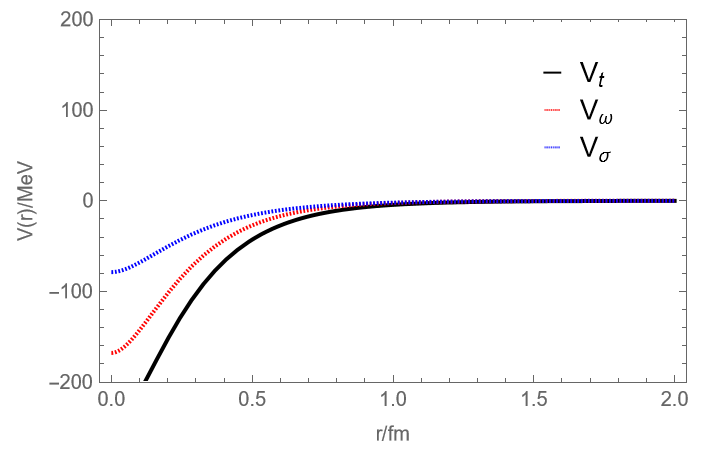}
    		\label{subfig:22}
    		\caption{}
    	\end{subfigure}
    	\hfill
    	\begin{subfigure}{0.32\textwidth}
    		\centering
    		\includegraphics[scale=0.4]{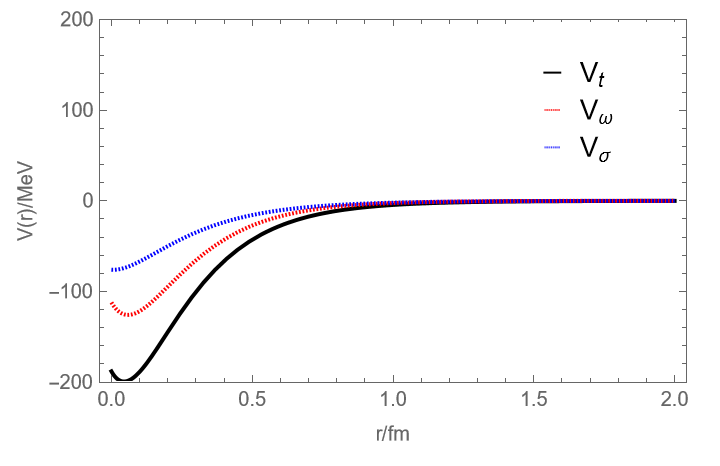}
    		\label{subfig:23}
    		\caption{}
    	\end{subfigure}
    	\hfill
    	\begin{subfigure}{0.32\textwidth}
    		\centering
    		\includegraphics[scale=0.4]{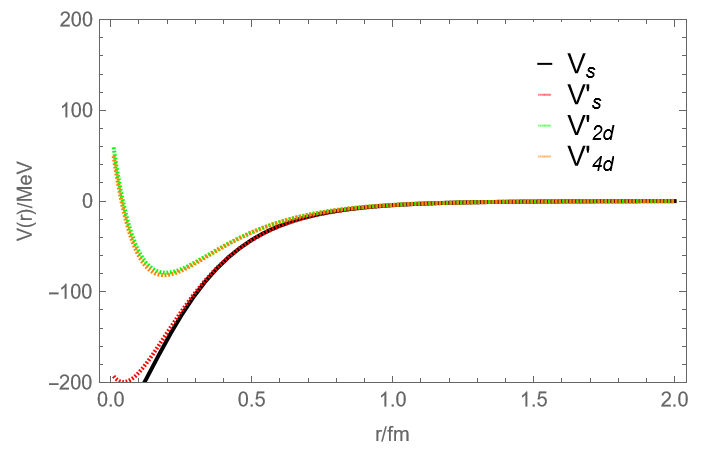}
    		\label{subfig:24}
    		\caption{}
    	\end{subfigure}
        \captionsetup{format=plain, justification=raggedright, singlelinecheck=false}
        \caption{The effective potentials for the $\Lambda_{b}\bar{B}^{*}$ system in the $\frac{1}{2}(\frac{1}{2}^{-})$ (upper row) and $\frac{1}{2}(\frac{3}{2}^{-})$  (lower row) channels. Panels (a,d): contributions from $\sigma$ and $\omega$ exchange without recoil corrections. Panels (b,e): same with recoil corrections. Panels (c,f): comparison of the total potentials in the ground state with and without recoil corrections. The cutoff value is $\Lambda=1.70$ GeV.}
    	\label{fig:6}
    \end{figure}

	The effective potentials for the $\frac{1}{2}(\frac{1}{2}^{-})$ channel of the $\Lambda_{b}\bar{B}$ system are presented in FIG. \ref{fig:5}. Panels (a) and (b) are the same as in FIG. \ref{fig:5}. The effective potentials for channels of the $\Lambda_{b}\bar{B}^{*}$ system are presented in FIG. \ref{fig:6}, where the upper and lower rows correspond to the $\frac{1}{2}(\frac{1}{2}^{-})$ and $\frac{1}{2}(\frac{3}{2}^{-})$ channels, respectively. Panels (a) to (f) are the same as in FIG. \ref{fig:2}.
	\par 
	As is shown in TABLE \ref{tab:7}, FIG. \ref{fig:5} and FIG. \ref{fig:6}, the exchange potentials from both $\sigma$ and $\omega$ are attractive. For the $\Lambda_{b}\bar{B}$ system, the impact of recoil corrections is minimal and almost negligible. In contrast, for the $\Lambda_{b}\bar{B}^{*}$ system, such corrections further weaken the binding, making the molecular state looser. 
	 \begin{table*}[htbp]
		\scriptsize
		\begin{center}
			\caption{\label{tab:7} The numerical results of the  $\Lambda_{b}\bar{B}$ and $\Lambda_{b}\bar{B}^{*}$ systems in different channels}
			\begin{tabular}{l ccc|cccccc}\toprule[1pt]
				\multicolumn{10}{c}{$\Lambda_{b}\bar{B}$ [$I(J^P)=\frac{1}{2}(\frac{1}{2}^-)$]}\\
				\midrule[1pt]
				\multicolumn{4}{c|}{Without recoil corrections} & \multicolumn{6}{c}{With recoil corrections}\\
				\midrule[1pt]
				$\Lambda$(GeV) & B.E.(MeV) & RMS(fm)  & $^2S_{\frac{1}{2}} $($\%$)  & $\Lambda$(GeV) & B.E.(MeV) & RMS(fm)  & $^2S_{\frac{1}{2}} $($\%$) & &  \\  
				1.30 & 0.55 & 2.87 & 100 & 1.30 & 0.55 & 2.88 & 100 & & \\
				1.40 & 2.28 & 1.56 & 100 & 1.40 & 2.27 & 1.56 & 100 & & \\
				1.50 & 4.97 & 1.14 & 100 & 1.50 & 4.95 & 1.14 & 100 & & \\	
				1.60 & 8.41 & 0.93 & 100 & 1.60 & 8.37 & 0.93 & 100 & & \\
				1.70 & 12.4 & 0.80 & 100 & 1.70 & 12.4 & 0.80 & 100 & & \\ 
				\midrule[1pt]
				\multicolumn{10}{c}{$\Lambda_{b}\bar{B}^{*}$ [$I(J^P)=\frac{1}{2}(\frac{1}{2}^-)$]}\\
				\midrule[1pt]
				\multicolumn{4}{c|}{Without recoil corrections} & \multicolumn{6}{c}{With recoil corrections}\\
				\midrule[1pt]
				$\Lambda$(GeV) & B.E.(MeV) & RMS(fm)  & $^2S_{\frac{1}{2}} $($\%$)  & $\Lambda$(GeV) & B.E.(MeV) & RMS(fm)  & $^2S_{\frac{1}{2}} $($\%$) &  $^4D_{\frac{1}{2}} $($\%$) & \\  
				1.30 & 0.58 & 2.81 & 100 & 1.30 & 0.49 & 3.01 & 100.0 & 0.00  &\\
				1.40 & 2.34 & 1.54 & 100 & 1.40 & 2.07 & 1.62 & 100.0 & 0.00  &\\
				1.50 & 5.05 & 1.13 & 100 & 1.50 & 4.52 & 1.19 & 99.99 & 0.01  &\\
				1.60 & 8.53 & 0.92 & 100 & 1.60 & 7.62 & 0.97 & 99.99 & 0.01  &\\
				1.70 & 12.6 & 0.80 & 100 & 1.70 & 11.2 & 0.84 & 99.98 & 0.02  &\\
				\midrule[1pt]
				\multicolumn{10}{c}{$\Lambda_{b}\bar{B}^{*}$ [$I(J^P)=\frac{1}{2}(\frac{3}{2}^-)$]}\\
				\midrule[1pt]
				\multicolumn{4}{c|}{Without recoil corrections} & \multicolumn{6}{c}{With recoil corrections}\\
				\midrule[1pt]
				$\Lambda$(GeV) & B.E.(MeV) & RMS(fm)  & $^4S_{\frac{3}{2}} $($\%$)  & $\Lambda$(GeV) & B.E.(MeV) & RMS(fm)  & $^4S_{\frac{3}{2}} $($\%$) & $^2D_{\frac{3}{2}} $($\%$) & $^4D_{\frac{3}{2}} $($\%$)   \\  
				1.30 & 0.58 & 2.81 & 100 & 1.30 & 0.49 & 3.01 & 100.0 & 0.00 & 0.00 \\
				1.40 & 2.34 & 1.54 & 100 & 1.40 & 2.07 & 1.62 & 100.0 & 0.00 & 0.00 \\
				1.50 & 5.05 & 1.13 & 100 & 1.50 & 4.52 & 1.19 & 99.99 & 0.00 & 0.01 \\
				1.60 & 8.53 & 0.92 & 100 & 1.60 & 7.62 & 0.97 & 99.99 & 0.00 & 0.01 \\
				1.70 & 12.6 & 0.80 & 100 & 1.70 & 11.2 & 0.84 & 99.98 & 0.01 & 0.01 \\		
				\bottomrule[1pt]				
			\end{tabular}
		\end{center}
	\end{table*}
	\clearpage
   
    \clearpage
    
    \section{Summary and Discussion} \label{sec4}
    Within the framework of the OBE model, we systematically study the $\Xi_{c}\bar{D}^{(*)}$, $\Lambda_{c}\bar{D}^{(*)}$, $\Xi_{c}D^{(*)}$ and $\Lambda_{c}D^{(*)}$ systems, along with their bottom analogues. We mainly focus on how recoil corrections affects the mass spectra of them. In addition, the $S$\--{}$D$ wave mixing effect is included. The recoil corrections are embodied in the effective potentials as momentum-related terms in the next and next-to-next leading order, including spin-spin interaction terms, spin-orbit force terms, and tensor force terms. Among them, the tensor force terms would not exist unless the recoil corrections are considered. The numerical results indicate that the recoil corrections, whose importance are always underestimated in previous research, can be crucial to the molecular pentaquark candidates.
    \par 
    The channels of the charmed systems and the bottomed systems share the same quantum numbers $I(J^{P})$. For the $\Xi_{c}\bar{D}$ system, channels with $I(J^P)=0(\frac{1}{2}^{-}), 1(\frac{1}{2}^{-})$ are allowed; for the $\Lambda_{c}\bar{D}$ system, only the channel with $I(J^P)=\frac{1}{2}(\frac{1}{2}^{-})$ is accessible; for the $\Xi_{c}\bar{D}^{*}$ system, channels with $I(J^P)=0(\frac{1}{2}^{-}), 0(\frac{3}{2}^{-}), 1(\frac{1}{2}^{-})$, $1(\frac{3}{2}^{-})$ are possible; for the $\Lambda_{c}\bar{D}^{*}$ system, channels with $I(J^P)=\frac{1}{2}(\frac{1}{2}^{-})$, $\frac{1}{2}(\frac{3}{2}^{-})$  are supported.       
    \par 
    After solving the coupled channel Schr{\"o}dinger equation, we cannot find a bound state solution in the $I=1$ channel of any systems. Besides, we also fail to discover a sign of molecular state in any channels of $\Lambda_{c}\bar{D}^{(*)}$ and $\Lambda_{c}D^{(*)}$ systems. Other channels possess bound state solutions whether the recoil corrections are taken into account.    
    \par
    For the $I(J^P)=0(\frac{1}{2}^{-})$ channel of $\Xi_{c}\bar{D}$ and $\Xi_{c}D$ systems, the $I(J^P)=0(\frac{1}{2}^{-})$ and $0(\frac{3}{2}^{-})$ channels of $\Xi_{c}\bar{D}^{*}$ and $\Xi_{c}D^{*}$ systems, as well as their bottom counterparts, and for the $I(J^P)=\frac{1}{2}(\frac{1}{2}^{-})$ and $\frac{1}{2}(\frac{3}{2}^{-})$ channels of $\Lambda_{b}\bar{B}^{*}$ system, the recoil corrections are unfavorable for the formation of bound states. Particularly, with a fixed cutoff value, the binding energy nearly halved in the $\Xi_{c}\bar{D}^{*}$ and $\Xi_{c}D^{*}$ systems.
    \par
    Based on the results presented above, we conclude three key findings and elaborate on them in detail. First, compared with the charmed systems, bound state solutions in the bottom systems are more readily obtained, and recoil corrections play a less significant role. This is attributed to the larger mass of the bottom systems, which decreases the kinetic energies and weakens the relativistic effects. Second, open-flavor systems are more likely to form bound states than their hidden-flavor counterparts, since the change in isospin factors turns the repulsive $\omega$ exchange potentials into attractive ones. Last and most importantly, the recoil corrections of the isospin singlet $\Xi_{c}\bar{D}^{*}$ and $\Xi_{c}D^{*}$ systems are far more striking than the others systems. To explain this phenomenon, we call back to the formulas of the effective potentials in Eq. (\ref{eq:21})\--(\ref{eq:24}) and the isospin factors in TABLE \ref{tab:4}. From the former, we find that the recoil corrections of the systems containing a $D$ meson emerge at $\mathcal{O}(\frac{1}{M^{2}})$, while such corrections of the systems containing a $D^{*}$ meson emerge at $\mathcal{O}(\frac{1}{M})$---an order of magnitude larger. From the latter, we observe that the absolute value of $C_{\rho}$ in these systems is three times that of $C_{\omega}$. Consequently, whether the recoil corrections from $\omega$ exchange counteract or reinforce those from $\rho$ exchange, the net diminution of the short-range attractive $\rho$ exchange potential from the $\mathcal{O}(\frac{1}{M})$ recoil corrections dominates the linear change of the $S$ wave.  
    \par   
    In conclusion, this research explores the double heavy and hidden heavy molecular pentaquark states with an SU(3) anti\--triplet baryon, which exhibits the great importance of the recoil corrections in the molecular candidates and further explain how they work. Consequently, our study holds significant implications which shed light on future investigations. From the experimental perspective, the molecular candidates predicted by this work are waiting for experimental evidence. From the theoretical perspective, more attention and in-depth studies of recoils corrections are needed so that people can gain a clearer understanding of the mechanism and dynamical effects of these higher-order corrections.    
    \clearpage

    \section{Acknowledgment} \label{sec5}
    
    This work is supported by the Fundamental Research Funds for the Central Universities under Grant No. 2022JBMC037 and the National Natural Science Foundation of China under Grants No. 12105009.

	\clearpage
	
    \section{Appendix}
    \subsection{Fourier transformations of \texorpdfstring{$\boldsymbol{q}$}{bold q} and \texorpdfstring{$\boldsymbol{k}$}{bold k} related structures}
    \vspace{-8pt}
    \begin{align*}
    	\frac{1}{-\boldsymbol{q}^2-m^2+i\epsilon}&=-\frac{m}{4\pi}H_{0},\\
    	\frac{\boldsymbol{q}^2}{-\boldsymbol{q}^2-m^2+i\epsilon}&=-\frac{m^{3}}{4\pi}H_{1},\\
    	\frac{q_i q_j}{-\boldsymbol{q}^2-m^2+i\epsilon}&=\left(\frac{m^3}{12\pi}\right)\left[H_3\left(\frac{3r_i r_j}{r^2}-\delta _{ij}  \right)-H_1\delta _{ij}\right],\\
    	\frac{i\left[\boldsymbol{\sigma}\cdot \left(\boldsymbol{q}\times \boldsymbol{k}\right)\right]}{-\boldsymbol{q}^{2}-m^{2}+i\epsilon}&=\frac{m^{3}}{4\pi}H_{2}\left(\boldsymbol{\sigma}\cdot \boldsymbol{L}\right),\\
    	\frac{\boldsymbol{k}^2}{-\boldsymbol{q}^2-m^2+i\epsilon}&=\frac{m}{4\pi}H_0 \boldsymbol{\nabla} ^2-\frac{m^3}{16\pi}H_1+\frac{m^3}{4\pi}H_2 \left(\boldsymbol{r} \cdot \boldsymbol{\nabla} \right), 
    \end{align*}
    where
    \begin{eqnarray*}
    	H_{0}&=&\frac{1}{m r}\left(e^{-m r}-e^{-\Lambda r}\right)-\frac{\Lambda^{2}-m^{2}}{2 m \Lambda} e^{-\Lambda r}, \\
    	H_{1}&=&-\frac{1}{m r}\left(e^{-m r}-e^{-\Lambda r}\right)+\Lambda \frac{\Lambda^{2}-m^{2}}{2 m^{2}} e^{-\Lambda r}, \\
    	H_{2}&=&-(1+m r) \frac{1}{m^{3} r^{3}} e^{-m r}+(1+\Lambda r) \frac{1}{m^{3} r^{3}} e^{-\Lambda r}+\frac{\Lambda^{2}-m^{2}}{2 m^{2}} \frac{1}{m r} e^{-\Lambda r}, \\
    	H_{3}&=&\left(1+\frac{3}{m r}+\frac{3}{m^{2} r^{2}}\right) \frac{1}{m r} e^{-m r}-\left(1+\frac{3}{\Lambda r}+\frac{3}{\Lambda^{2} r^{2}}\right) \frac{\Lambda^{2}}{m^{2}} \frac{1}{m r} e^{-\Lambda r}-\frac{\Lambda^{2}-m^{2}}{2 m^{2}}(1+\Lambda r) \frac{1}{m r} e^{-\Lambda r}.
    \end{eqnarray*}
	\subsection{Representations of momentum-related operators}\label{subsecI}
	\begin{table*}[!htbp]
		\scriptsize
		\begin{center}
			\caption{\label{tab:8}  The matrix elements $\left\langle f \mid O_{k} \mid i \right\rangle$ for the angular momentum related operators $O_{k}$  }
			\begin{tabular}{c|c|c|c|c|c|c}
				\toprule[1pt]
				& $O_{1}$ & $O_{2}$ & $O_{3}$ & $O_{4}$ & $O_{5}$ & $O_{6}$ \\
				\midrule[1pt]
				$\left\langle f \mid O_{k} \mid i \right\rangle_{[\frac{1}{2}]}$ 
				& $\operatorname{Diag}(1,1)$ 
				& $\begin{pmatrix} 0 & -\sqrt{2} \\ -\sqrt{2} & 1 \end{pmatrix}$ 
				& $\begin{pmatrix} 0 & -\sqrt{2} \\ -\sqrt{2} & -2 \end{pmatrix}$ 
				& $\operatorname{Diag}(0,-3)$ 
				& $\operatorname{Diag}(-2,1)$ 
				& $\operatorname{Diag}(0,-3)$ \\
				\midrule
				$\left\langle f \mid O_{k} \mid i \right\rangle_{[\frac{3}{2}]}$ 
				& $\operatorname{Diag}(1,1,1)$ 
				& $\begin{pmatrix} 0 & 1 & -1 \\ 1 & 0 & -1 \\ -1 & -1 & 0 \end{pmatrix}$ 
				& $\begin{pmatrix} 0 & 1 & 2 \\ 1 & 0 & -1 \\ 2 & -1 & 0 \end{pmatrix}$ 
				& $\begin{pmatrix} 0 & 0 & 0 \\ 0 & -2 & 1 \\ 0 & 1 & -2 \end{pmatrix}$ 
				& $\operatorname{Diag}(1,-2,1)$ 
				& $\begin{pmatrix} 0 & 0 & 0 \\ 0 & 1 & -2 \\ 0 & -2 & -2 \end{pmatrix}$ \\
				\bottomrule[1pt]
			\end{tabular}
		\end{center}
	\end{table*}
	\par 
	The operators $O_1$,$O_2$ $\dots$ $O_5$ in TABLE \ref{tab:8} read
	    \begin{align*}
		O_{1}&=\boldsymbol{\epsilon}_2 \cdot \boldsymbol{\epsilon}_4^\dagger,
		\\
		O_{2}&=S(\hat{\boldsymbol{r}}, \boldsymbol{\epsilon}_2, \boldsymbol{\epsilon}_4^\dagger),
		\\
		O_{3}&=S(\hat{\boldsymbol{r}}, i\boldsymbol{\epsilon}_2 \cdot \boldsymbol{\epsilon}_4^\dagger, \boldsymbol{\sigma}),
		\\   O_{4}&=i(\boldsymbol{\epsilon}_2 \times \boldsymbol{\epsilon}_4^\dagger) \cdot \boldsymbol{L},
		\\    O_{5}&=i(\boldsymbol{\epsilon}_2 \times \boldsymbol{\epsilon}_4^\dagger) \cdot \boldsymbol{\sigma},
		\\    O_{6}&=(\boldsymbol{\epsilon}_2 \cdot \boldsymbol{\epsilon}_4^\dagger)(\boldsymbol{\sigma} \cdot \boldsymbol{L}),		 
	\end{align*} 
	where $S(\hat{\boldsymbol{r}}, \boldsymbol{a}, \boldsymbol{b})=3 \frac{(\boldsymbol{a} \cdot \boldsymbol{r})(\boldsymbol{b} \cdot \boldsymbol{r})}{r^{2}}-\boldsymbol{a} \cdot \boldsymbol{b}$, $\boldsymbol{L}=\boldsymbol{r} \times (-i\boldsymbol{\nabla})$.

	\subsection{The numerical results of $\Xi_{b}B^{(*)}$ and $\Xi_{b}\bar{B}^{(*)}$ systems}
	\begin{table}[!t]
		\scriptsize
		\begin{center}
			\caption{\label{tab:9} The numerical results of the  $\Xi_{b}B$, $\Xi_{c}B^{*}$, $\Xi_{b}\bar{B}$ and $\Xi_{c}\bar{B}^{*}$ systems in different channels}
			\begin{tabular}{l ccc|cccccc}\toprule[1pt]
				\multicolumn{10}{c}{$\Xi_{b}B$ [$I(J^P)=0(\frac{1}{2}^-)$] }\\
				\midrule[1pt]
				\multicolumn{4}{c|}{Without recoil corrections} & \multicolumn{6}{c}{With recoil corrections}\\
				\midrule[1pt]
				$\Lambda$(GeV) & B.E.(MeV) & RMS(fm)  & $^2S_{\frac{1}{2}} $($\%$)  & $\Lambda$(GeV) & B.E.(MeV) & RMS(fm)  & $^2S_{\frac{1}{2}} $($\%$) & &  \\  
		    	1.10 & 0.54 & 2.89 & 100 & 1.10 & 0.54 & 2.90 & 100 && \\
			    1.15 & 2.08 & 1.64 & 100 & 1.15 & 2.07 & 1.64 & 100 && \\
			    1.20 & 4.51 & 1.21 & 100 & 1.20 & 4.49 & 1.21 & 100 && \\
			    1.25 & 7.75 & 0.98 & 100 & 1.25 & 7.71 & 0.99 & 100 && \\
			    1.30 & 11.7 & 0.85 & 100 & 1.30 & 11.6 & 0.85 & 100 && \\
				\midrule[1pt]
				\multicolumn{10}{c}{$\Xi_{c}B^{*}$ [$I(J^P)=0(\frac{1}{2}^-)$] }\\
				\midrule[1pt]
				\multicolumn{4}{c|}{Without recoil corrections} & \multicolumn{6}{c}{With recoil corrections}\\
				\midrule[1pt]
				$\Lambda$(GeV) & B.E.(MeV) & RMS(fm)  & $^2S_{\frac{1}{2}} $($\%$)  & $\Lambda$(GeV) & B.E.(MeV) & RMS(fm)  & $^2S_{\frac{1}{2}} $($\%$) & $^4D_{\frac{1}{2}} $($\%$) & \\  
				1.10 & 0.57 & 2.83 & 100 & 1.10 & 0.49 & 3.03 & 100.0 & 0.00 &\\
				1.15 & 2.13 & 1.62 & 100 & 1.15 & 1.90 & 1.70 & 100.0 & 0.00 &\\
				1.20 & 4.59 & 1.20 & 100 & 1.20 & 4.14 & 1.25 & 99.99 & 0.01 &\\
				1.25 & 7.85 & 0.98 & 100 & 1.25 & 7.12 & 1.02 & 99.99 & 0.01 &\\
				1.30 & 11.8 & 0.84 & 100 & 1.30 & 10.7 & 0.88 & 99.99 & 0.01 &\\
				\midrule[1pt]
				\multicolumn{10}{c}{$\Xi_{c}B^{*}$ [$I(J^P)=0(\frac{3}{2}^-)$] }\\
				\midrule[1pt]
				\multicolumn{4}{c|}{Without recoil corrections} & \multicolumn{6}{c}{With recoil corrections}\\
				\midrule[1pt]
				$\Lambda$(GeV) & B.E.(MeV) & RMS(fm)  & $^4S_{\frac{3}{2}} $($\%$)  & $\Lambda$(GeV) & B.E.(MeV) & RMS(fm)  & $^4S_{\frac{3}{2}} $($\%$) & $^2D_{\frac{3}{2}} $($\%$) & $^4D_{\frac{3}{2}} $($\%$) \\  
				1.10 & 0.57 & 2.83 & 100 & 1.10 & 0.49 & 3.03 & 100.0 & 0.00 & 0.00 \\
				1.15 & 2.13 & 1.62 & 100 & 1.15 & 1.90 & 1.70 & 100.0 & 0.00 & 0.00 \\
				1.20 & 4.59 & 1.20 & 100 & 1.20 & 4.14 & 1.25 & 99.99 & 0.00 & 0.01 \\
				1.25 & 7.85 & 0.98 & 100 & 1.25 & 7.12 & 1.02 & 99.99 & 0.00 & 0.01 \\
				1.30 & 11.8 & 0.84 & 100 & 1.30 & 10.7 & 0.88 & 99.99 & 0.00 & 0.01 \\
				\midrule[1pt]
				\multicolumn{10}{c}{$\Xi_{b}\bar{B}$ [$I(J^P)=0(\frac{1}{2}^-)$] }\\
				\midrule[1pt]
				\multicolumn{4}{c|}{Without recoil corrections} & \multicolumn{6}{c}{With recoil corrections}\\
				\midrule[1pt]
				$\Lambda$(GeV) & B.E.(MeV) & RMS(fm)  & $^2S_{\frac{1}{2}} $($\%$)  & $\Lambda$(GeV) & B.E.(MeV) & RMS(fm)  & $^2S_{\frac{1}{2}} $($\%$) & &  \\  
				1.00 & 0.69 & 2.63 & 100 & 1.00 & 0.68 & 2.63 & 100 && \\
				1.05 & 3.96 & 1.29 & 100 & 1.05 & 3.94 & 1.29 & 100 && \\
				1.10 & 9.76 & 0.92 & 100 & 1.10 & 9.71 & 0.93 & 100 && \\
				1.15 & 17.8 & 0.75 & 100 & 1.15 & 17.7 & 0.75 & 100 && \\
				1.20 & 27.9 & 0.64 & 100 & 1.20 & 27.8 & 0.64 & 100 && \\
				\midrule[1pt]
				\multicolumn{10}{c}{$\Xi_{c}\bar{B}^{*}$ [$I(J^P)=0(\frac{1}{2}^-)$] }\\
				\midrule[1pt]
				\multicolumn{4}{c|}{Without recoil corrections} & \multicolumn{6}{c}{With recoil corrections}\\
				\midrule[1pt]
				$\Lambda$(GeV) & B.E.(MeV) & RMS(fm)  & $^2S_{\frac{1}{2}} $($\%$)  & $\Lambda$(GeV) & B.E.(MeV) & RMS(fm)  & $^2S_{\frac{1}{2}} $($\%$) & $^4D_{\frac{1}{2}} $($\%$) & \\  
				1.00 & 0.71 & 2.58 & 100 & 1.00 & 0.62 & 2.75 & 100.0 & 0.00 &\\
				1.05 & 4.03 & 1.28 & 100 & 1.05 & 3.66 & 1.33 & 99.99 & 0.01 &\\
				1.10 & 9.87 & 0.92 & 100 & 1.10 & 9.02 & 0.95 & 99.99 & 0.01 &\\
				1.15 & 18.0 & 0.74 & 100 & 1.15 & 16.4 & 0.77 & 99.98 & 0.02 &\\
				1.20 & 28.1 & 0.64 & 100 & 1.20 & 25.6 & 0.67 & 99.97 & 0.03 &\\
				\midrule[1pt]
				\multicolumn{10}{c}{$\Xi_{c}\bar{B}^{*}$ [$I(J^P)=0(\frac{3}{2}^-)$] }\\
				\midrule[1pt]
				\multicolumn{4}{c|}{Without recoil corrections} & \multicolumn{6}{c}{With recoil corrections}\\
				\midrule[1pt]
				$\Lambda$(GeV) & B.E.(MeV) & RMS(fm)  & $^4S_{\frac{3}{2}} $($\%$)  & $\Lambda$(GeV) & B.E.(MeV) & RMS(fm)  & $^4S_{\frac{3}{2}} $($\%$) & $^2D_{\frac{3}{2}} $($\%$) & $^4D_{\frac{3}{2}} $($\%$) \\  
				1.00 & 0.71 & 2.58 & 100 & 1.00 & 0.62 & 2.75 & 100.0 & 0.00 & 0.00 \\
				1.05 & 4.03 & 1.28 & 100 & 1.05 & 3.66 & 1.33 & 99.99 & 0.00 & 0.01 \\
				1.10 & 9.87 & 0.92 & 100 & 1.10 & 9.02 & 0.95 & 99.99 & 0.00 & 0.01 \\
				1.15 & 18.0 & 0.74 & 100 & 1.15 & 16.4 & 0.77 & 99.98 & 0.01 & 0.01 \\
				1.20 & 28.1 & 0.64 & 100 & 1.20 & 25.6 & 0.67 & 99.97 & 0.01 & 0.02 \\
				\bottomrule[1pt]								
			\end{tabular}
		\end{center}
	\end{table}


\clearpage
    
    \begin{thebibliography}{99}  
    	\bibitem{ref1}
    	M.~Gell-Mann,
    	Phys. Lett. \textbf{8}, 214-215 (1964).
    	
    	\bibitem{ref2}
    	G.~Zweig,
        An SU(3) model for strong interaction symmetry and its breaking. Version 1, (1964).
    	
    	\bibitem{ref3}
    	S.~K.~Choi \textit{et al.} [Belle],
    	Phys. Rev. Lett. \textbf{91}, 262001 (2003).
    	
    	
    	\bibitem{ref4}
    	K.~Abe \textit{et al.} [Belle],
    	Phys. Rev. Lett. \textbf{94}, 182002 (2005).
    	
    	\bibitem{ref5}
    	B.~Aubert \textit{et al.} [BaBar],
    	Phys. Rev. Lett. \textbf{101}, 082001 (2008).
    	
    	\bibitem{ref6}
    	A.~Bondar \textit{et al.} [Belle],
    	Phys. Rev. Lett. \textbf{108}, 122001 (2012).
    	
    	\bibitem{ref7}
    	M.~Ablikim \textit{et al.} [BESIII],
    	Phys. Rev. Lett. \textbf{110}, 252001 (2013).
    	
    	\bibitem{ref8}
    	R.~Aaij \textit{et al.} [LHCb],
    	Phys. Rev. Lett. \textbf{118}, no.2, 022003 (2017).
    	
    	\bibitem{ref9}
    	R.~Aaij \textit{et al.} [LHCb],
    	Phys. Rev. Lett. \textbf{115}, 072001 (2015).
    	
    	\bibitem{ref10}
    	R.~Aaij \textit{et al.} [LHCb],
    	Phys. Rev. Lett. \textbf{122}, no.22, 222001 (2019).
    	
    	\bibitem{ref11}
    	R.~Aaij \textit{et al.} [LHCb],
    	Nature Phys. \textbf{18}, no.7, 751-754 (2022).
    	
    	\bibitem{ref12}
    	X.~Liu, Z.~G.~Luo, Y.~R.~Liu and S.~L.~Zhu,
    	Eur. Phys. J. C \textbf{61}, 411-428 (2009).
    	
    	\bibitem{ref13}
    	Y.~R.~Liu, M.~Oka, M.~Takizawa, X.~Liu, W.~Z.~Deng and S.~L.~Zhu,
    	Phys. Rev. D \textbf{82}, 014011 (2010).
    	
    	\bibitem{ref14}
    	Z.~F.~Sun, J.~He, X.~Liu, Z.~G.~Luo and S.~L.~Zhu,
    	Phys. Rev. D \textbf{84}, 054002 (2011).
    	
    	\bibitem{ref15}
    	L.~Roca, J.~Nieves and E.~Oset,
    	Phys. Rev. D \textbf{92}, no.9, 094003 (2015).
    	
    	\bibitem{ref16}
        H.~Hogaasen, J.~M.~Richard and P.~Sorba,
        Phys. Rev. D \textbf{73}, 054013 (2006).
    	
    	\bibitem{ref17}
    	Y.~Cui, X.~L.~Chen, W.~Z.~Deng and S.~L.~Zhu,
    	HEPNP \textbf{31}, 7-13 (2007).
    	
    	\bibitem{ref18}
    	E.~Santopinto and A.~Giachino,
    	Phys. Rev. D \textbf{96}, no.1, 014014 (2017).
    	
    	\bibitem{ref19}
    	B.~A.~Li,
    	Phys. Lett. B \textbf{605}, 306-310 (2005).
    	
    	\bibitem{ref20}
    	S.~L.~Zhu,
    	Phys. Lett. B \textbf{625}, 212 (2005).
    	
    	\bibitem{ref21}
    	C.~A.~Meyer and E.~S.~Swanson,
    	Prog. Part. Nucl. Phys. \textbf{82}, 21-58 (2015).
    	
    	\bibitem{ref22}
    	Y.~M.~Cho, X.~Y.~Pham, P.~Zhang, J.~J.~Xie and L.~P.~Zou,
    	Phys. Rev. D \textbf{91}, no.11, 114020 (2015).
    	
    	\bibitem{ref23}
    	C.~J.~Morningstar and M.~J.~Peardon,
    	Phys. Rev. D \textbf{60}, 034509 (1999).
    	
    	\bibitem{ref24}
        M.~Loan, X.~Q.~Luo and Z.~H.~Luo,
        Int. J. Mod. Phys. A \textbf{21}, 2905-2936 (2006).
    	
    	\bibitem{ref25}
    	M.~Loan and Y.~Ying,
    	Prog. Theor. Phys. \textbf{116}, 169 (2006).
    	   	
    	\bibitem{ref26}
    	N.~A.~Tornqvist,
    	[arXiv:hep-ph/0308277 [hep-ph]], (2003). 
    	
    	\bibitem{ref27}
    	E.~S.~Swanson,
    	Phys. Lett. B \textbf{588}, 189-195 (2004).
    	
    	\bibitem{ref28}
    	N.~Li and S.~L.~Zhu,
    	Phys. Rev. D \textbf{86}, 074022 (2012).
    	
    	\bibitem{ref29}
    	G.~J.~Ding,
    	Phys. Rev. D \textbf{79}, 014001 (2009).
    	
    	\bibitem{ref30}
    	Q.~Wang, C.~Hanhart and Q.~Zhao,
    	Phys. Rev. Lett. \textbf{111}, no.13, 132003 (2013).
    	
    	\bibitem{ref31}
    	M.~Cleven, Q.~Wang, F.~K.~Guo, C.~Hanhart, U.~G.~Mei{\ss}ner and Q.~Zhao,
    	Phys. Rev. D \textbf{90}, no.7, 074039 (2014).
    	
    	\bibitem{ref32}
    	M.~L.~Du, V.~Baru, F.~K.~Guo, C.~Hanhart, U.~G.~Mei{\ss}ner, J.~A.~Oller and Q.~Wang,
    	Phys. Rev. Lett. \textbf{124}, no.7, 072001 (2020).
    	
    	\bibitem{ref33}
    	N.~Li, Z.~F.~Sun, X.~Liu and S.~L.~Zhu,
    	Chin. Phys. Lett. \textbf{38}, no.9, 092001 (2021).
    	
    	\bibitem{ref34}
    	N.~Li, Z.~F.~Sun, X.~Liu and S.~L.~Zhu,
    	Phys. Rev. D \textbf{88}, no.11, 114008 (2013).
    	
    	\bibitem{ref35}
    	H.~Yukawa,
    	Proc. Phys. Math. Soc. Jap. \textbf{17}, 48-57 (1935).
    	
    	\bibitem{ref36}
    	X.~Chen and L.~Ma,
    	Eur. Phys. J. A \textbf{61}, no.6, 150 (2025).
    	
    	\bibitem{ref37}
    	R.~Aaij \textit{et al.} [LHCb],
    	Phys. Rev. Lett. \textbf{131}, no.3, 031901 (2023).
    	
    	\bibitem{ref38}
        F.~L.~Wang and X.~Liu,
        Phys. Lett. B \textbf{835}, 137583 (2022).
    	
    	\bibitem{ref39}
    	H.~X.~Chen, W.~Chen, X.~Liu and X.~H.~Liu,
    	Eur. Phys. J. C \textbf{81}, no.5, 409 (2021).
    	
    	\bibitem{ref40}
    	H.~W.~Ke, F.~Lu, H.~Pang, X.~H.~Liu and X.~Q.~Li,
    	Eur. Phys. J. C \textbf{83}, no.11, 1074 (2023).
    	
    	\bibitem{ref41}
    	A.~Feijoo, W.~F.~Wang, C.~W.~Xiao, J.~J.~Wu, E.~Oset, J.~Nieves and B.~S.~Zou,
    	Phys. Lett. B \textbf{839}, 137760 (2023).
    	
    	\bibitem{ref42}
    	B.~Wang, K.~Chen, L.~Meng and S.~L.~Zhu,
    	Phys. Rev. D \textbf{110}, no.1, 014038 (2024).
    	
    	\bibitem{ref43}
    	N.~Lee, Z.~G.~Luo, X.~L.~Chen and S.~L.~Zhu,
    	Phys. Rev. D \textbf{84}, 014031 (2011).
    	
    	\bibitem{ref44}
    	S.~Navas \textit{et al.} [Particle Data Group],
    	Phys. Rev. D \textbf{110}, no.3, 030001 (2024).
    	
    	\bibitem{ref45}
    	N.~Li and S.~L.~Zhu,
    	Phys. Rev. D \textbf{86}, 014020 (2012).
    	
    	\bibitem{ref46}
    	L.~Zhao, L.~Ma and S.~L.~Zhu,
    	Phys. Rev. D \textbf{89}, no.9, 094026 (2014).
    	
    	\bibitem{ref47}
    	F.~L.~Wang, R.~Chen and X.~Liu,
    	Phys. Lett. B \textbf{835}, 137502 (2022).
    	
    	\bibitem{ref48}
    	X.~K.~Dong, F.~K.~Guo and B.~S.~Zou,
    	Progr. Phys. \textbf{41}, 65-93 (2021).
    	
    	\bibitem{ref49}
    	E.~Klempt, F.~Bradamante, A.~Martin and J.~M.~Richard,
    	Phys. Rept. \textbf{368}, 119-316 (2002).  	
    \end{thebibliography}
\end{document}